\documentclass[12pt]{article}
\usepackage[left=1.25in,right=1.25in,top=1in,bottom=1in]{geometry}
\usepackage{authblk}
\usepackage{microtype}
\usepackage{ragged2e}
\justifying

\usepackage[round]{natbib}
\usepackage{xcolor}

\usepackage{color,hyperref}
\definecolor{darkblue}{rgb}{0.1,0.1,0.8}
\hypersetup{colorlinks,breaklinks,
            linkcolor=darkblue,urlcolor=darkblue,
            anchorcolor=darkblue,citecolor=darkblue}

\usepackage{graphicx}
\usepackage{amsmath,amsbsy,amssymb,bbm}
\usepackage{multirow,multicol,subfigure,booktabs}
\usepackage{natbib}
\usepackage{algorithm}
\usepackage[noend]{algpseudocode}
\newcommand{\E}{\mathop{\mathbb E}}

\providecommand{\keywords}[1]{\small \textbf{Keywords:} #1}

\begin{document}

\title{Conditional copula graphic estimator for semi-competing risks data}
\date{}

\author[1]{\small Shamsia Sobhan}
\author[2, 3, 4,*]{\small Elif Fidan Acar}

\affil[1]{\footnotesize Children's Hospital Research Institute of Manitoba, University of Manitoba, Winnipeg, Canada}

\affil[2]{\footnotesize Department of Mathematics and Statistics, University of Guelph, Guelph, Canada}
\affil[3]{\footnotesize Hospital for Sick Children, Toronto, Canada}
\affil[4]{\footnotesize Department of Statistics, University of Manitoba, Winnipeg, Canada}
\affil[*]{Corresponding author: Elif Fidan Acar, eacar@uoguelph.ca}
\maketitle

\begin{abstract}
In semi-competing risks data, the interest lies in the estimation of the survival function of a non-terminal event time, which is subject to dependent censoring by a terminal event. 
This problem has been extensively studied in the literature, but mostly focusing on unconditional settings. 
However, in many clinical applications incorporating covariates is necessary to control for confounding and improve survival function estimation.
In this paper, we propose a conditional copula-graphic estimator that allows for covariate adjustment in the marginal survival functions of the non-terminal and terminal event times as well as in their dependence structure.
The proposed estimator is semiparametric in that the conditional copula is specified parametrically using an Archimedean copula, but its dependence parameter function and margins are estimated nonparametrically. The estimator is obtained via a sequential iterative algorithm with alternating updates of the survival function of the non-terminal event and the conditional copula. 
The performance of the conditional copula-graphic estimator is assessed using simulated and real data, and is compared to that of the unconditional copula-graphic estimator to investigate the consequences of failing to account for covariate effects.
\end{abstract}

\keywords{
Beran's estimator, copula-graphic estimator, covariate adjustment, iterative algorithm, semi-competing risks, semiparametric estimation.}

\section{Introduction}
\label{s:intro}

Semi-competing risks refer to a situation when a subject may experience a terminal event (e.g., death) before the occurrence of a non-terminal event (e.g., cancer recurrence) where both events are subject to independent (administrative) censoring. 
Since the censoring of the non-terminal event by the terminal event is informative, the dependence between the two event times needs to be accounted for when estimating the marginal survival function of the non-terminal event time.

As an example, the lifetime and cancer recurrence time of a patient with cancer history are likely to be dependent as both events are subject to some common risk factors such as genetic background, immune system and patient's lifestyle. A patient must experience cancer recurrence necessarily prior to death; and both events can be independently censored by the end of the study. The statistical analysis for such data needs to account for the dependence structure of the lifetime and cancer recurrence time along with their censoring mechanism. If the terminal event is observed for all study subjects, the data structure falls under dependent right-censoring.

In the case of dependent right-censored data, the copula-graphic estimator \citep{zheng1995estimates} is commonly used to estimate the marginal survival function of the event time of interest. This estimator has a closed-form expression when the copula of the non-terminal and terminal event times is Archimedean \citep{rivest2001martingale}. For semi-competing risks data, where there is additional independent censoring, \cite{lakhal2008estimating} proposed a copula-graphic estimator using one parameter Archimedean copulas, which was extended to multi-parameter Archimedean copulas by \cite{heuchenne2014likelihood}. The large sample properties of the copula-graphic estimator were studied by \cite{laurent2013estimating} and \cite{rivest2001martingale}.

While semi-competing risks data have been extensively studied in the literature, only few works have addressed the incorporation of covariates in the modeling strategy. Most research in this domain focused on regression analysis of marginal survival functions; though some also considered potential covariate effects on the dependence structure for some special cases, such as discrete covariates \citep{hsieh2008regression, chen2012maximum, Wei:2023, Wang:2024} and time-varying dependence parameters \citep{peng2007regression, hsieh2012regression}. In some other works,  the dependence structure is specified using frailty models \citep{ghosh2006semiparametric, xu2010statistical}. Additional work includes quantile regression for analyzing semi-competing risks data in the presence of covariates  \citep{li2015quantile, yang2016new}. Despite these efforts, most existing approaches either depend on restrictive parametric assumptions or can only accommodate binary covariates, highlighting the need for more flexible regression frameworks for semi-competing risks.

This paper addresses this need by presenting a conditional copula-graphic estimator, which allows for covariate adjustment in both the marginal survival functions and the dependence structure of non-terminal and terminal event times. Our work extends the conditional copula-graphic estimator of \cite{braekers2005copula} for dependent right-censored data to the semi-competing risks data setting. Besides accounting for additional independent censoring arising in semi-competing risks data, a major contribution in our work is that, while \cite{braekers2005copula} assumed that the conditional copula is known, here we relax this assumption and propose an iterative estimation algorithm to sequentially estimate the conditional copula-graphic estimator and the conditional copula, essential for practical applications.  
Our work further extends the copula-graphic estimator of \cite{heuchenne2014likelihood} for the unconditional case to more realistic but challenging regression settings. 
Through extensive simulations, we evaluate the performance of the proposed estimator under different rates of dependent censoring and for different dependence structures, and demonstrate its utility in two data applications. 
We further outline a nonparametric bootstrap procedure to construct pointwise confidence intervals for the conditional survival function of the terminal event as well as for the dependence parameter linking the non-terminal and terminal event times.

In survival analysis, inclusion of covariates is necessary to control for confounding and improve survival function estimation. Clinicians and researchers routinely face the challenge of whether to include a covariate or not. 
Hence, of particular interest in this work is to assess the consequences of failing to account for covariate effects in the marginal survival functions and/or the dependence structure when analyzing semi-competing risks data. Considering situations where (i) there is no covariate effect, (ii) a covariate affects only the margins but not the dependence structure, and (iii) a covariate affects both the margins and the dependence structure, we provide a detailed assessment of the cost of ignoring covariate effects in estimation performance via comparisons of the conditional and unconditional copula graphic estimators.

The paper is organized as follows. Section~\ref{s:model} introduces the model and describes the proposed conditional copula-graphic estimator along with the iterative estimation algorithm. Section~\ref{s:simulation} presents the results from the simulation study comparing the conditional versus unconditional copula-graphic estimators.
Section~\ref{s:data} contains applications to data from the Stanford Heart Transplant Study \citep{clark1971cardiac} and the Bone Marrow Transplant Study \citep{klein2006survival}.
Section~\ref{s:conc} concludes with a brief discussion.

\section{Conditional Copula-Graphic Estimator}
\label{s:model}

This section introduces the joint model for the non-terminal and terminal event times, outlines estimation of the model components, and presents the proposed conditional copula-graphic estimator obtained through an iterative algorithm.

\subsection{Model}

Let $Y_1$ be the non-terminal event time, $Y_2$ be the terminal event time and $Z$ be the censoring time. 
The observed random variables are $T_1 = \min\{Y_1, Y_2, Z\}$, $T_2 = \min\{Y_2, Z\}$, $\Delta_1 = \mathbbm{1} \{Y_1 \le Y_2, Y_1 \le Z\}$ and $\Delta_2 = \mathbbm{1} \{Y_2 \le Z\}$. When $\Delta_1=\Delta_2=1$, both $Y_1$ and $Y_2$ are observed, when $\Delta_1=1$,  $Y_1$ and the minimum of $(Y_2, Z)$ are observed and when $\Delta_2=1$,  $Y_2$ is observed. The situation where either non-terminal or terminal or both events occur is defined using the indicator $\Delta_3 = \mathbbm{1}\{ \min(Y_1, Y_2) < Z\} = \min \{1, \Delta_1 + \Delta_2\}$ for $T^\ast = \min\{Y_1,Y_2\}$.

Let $X$ be a continuous covariate that affects both the marginal survival functions and the dependence structure of $Y_1$ and $Y_2$, and suppose that $(Y_1, Y_2)$ is conditionally independent of $Z$ given $X=x$.
Then, the conditional joint survival function of $(Y_1, Y_2)$ given $X=x$ can be represented as 
\begin{equation} 
\label{model}
H_X (t_1, t_2 \mid x) = \mathcal{C}_X \{ S_{1\mid X} (t_1\mid x), S_{2\mid X}(t_2\mid x) \mid x \}, \qquad t_1 \leq t_2
\end{equation}
where $S_{j\mid X}(t_j \mid x ) = P(Y_j > t_j  \mid X=x)$ is the conditional marginal survival function of $Y_j$ given $X=x$ for $j=1,2$, and $\mathcal{C}_X$ is the conditional copula that describes the conditional dependence between $Y_1$ and $Y_2$ given $X=x$.
We assume that, for each $x$ in the range of the covariate, $\mathcal{C}_X$ belongs to the same Archimedean copula family, but its dependence parameter is allowed to change with the value of the covariate. 
Hence, $\mathcal{C}_X$  takes the form
\begin{equation} 
\label{ccop}
\mathcal{C}_X\{S_{1\mid X} (t_1\mid x), S_{2\mid X}(t_2\mid x) \mid x\} 
= \psi_{\theta(x)}^{-1}  \left[   \psi_{\theta(x)} \{S_{1\mid X} (t_1\mid x) \}   + \psi_{\theta(x)}\{S_{2\mid X}(t_2\mid x)\} \right],
\end{equation}
where $\theta(x)$ is the copula parameter and $\psi_{\theta(x)}: [0, 1] \to [0,\infty]$ is the corresponding Archimedean generator, which is a continuous, strictly decreasing, convex function with $ \psi(0) = \infty$ and  $\psi(1) = 0$.
The generator functions  for some one-parameter Archimedean families are provided in Table~\ref{tab1}, along with their inverses and derivatives.
For one-parameter Archimedean families where  $\psi'_{\theta_1} / \psi'_{\theta_2}$ is strictly monotone for $\theta_1\neq \theta_2$,  \cite{heuchenne2014likelihood} showed the identifiability of the unconditional version of model in \eqref{model}. 
In the conditional case, the same results holds for each covariate value $x$. Hence, the model in \eqref{model} is identifiable.

Given the observed data $\{(T_{1i}, T_{2i}, \Delta_{1i}, \Delta_{2i}, x_i);  i = 1, \ldots, n\}$, one needs to estimate $S_{1\mid X}(\cdot \mid x)$, $S_{2\mid X}(\cdot \mid x)$ and $\theta(x)$ to fit the model in \eqref{model}, which we address in the following.

\begin{table}
\begin{center}
\caption{Archimedean copulas and their parameterizations, generator and inverse generator functions.}
\label{tab1}
\scalebox{0.65}{
\begin{tabular}{l c c c}
\hline
\\
&  Clayton & Frank & Gumbel
\\
\hline
\\ 
$ \mathcal{C}(u_1, u_2)$        
&      $ \displaystyle   (u_1^{-\theta} + u_2^{-\theta} -1)^{-\frac{1}{\theta}} $ 
&      $ \displaystyle -\frac{1}{\theta}\ln \left\{ 1 + \frac{(e^{-\theta u_1} - 1)(e^{-\theta u_2} - 1)}{e^{-\theta} - 1}\right\}$
&      $ \displaystyle  \exp\left[ - \{  (-\ln u_1)^\theta + (-\ln u_2)^\theta\}^{\frac{1}{\theta}} \right]$
\\
\\[1ex]
Parametrizations:
\\[1ex]
$\theta  \in  $       &   $(0, \infty$)   &   $(-\infty, \infty)/\{0\}$    &      $[1, \infty)$
\\[2ex]
$\tau$     &      $\displaystyle {\theta}/ {(\theta + 2)}$      &        \text{no closed form}     &      $\displaystyle 1 - 1/{\theta}$
\\[2ex]
$g^{-1}(\eta)$      &   $\exp(\eta)$    &     $\eta$    &     $\exp(\eta) + 1$
\\[3ex]
Generator Functions:
\\[1ex]
$\psi_\theta(z)$     & $ \displaystyle (z^{-\theta}-1) / \theta$ &  $\displaystyle -\ln \left( \frac{ e^{-\theta z} - 1}{e^{-\theta} - 1}\right) $ & $(-\ln z)^\theta$
\\[2ex]
$\psi_\theta'(z)$     &  $ \displaystyle -z^{-\theta - 1}$&       $\displaystyle  \theta e^{- \theta z}  /  (e^{- \theta z} - 1)$     & $ - \theta  (-\ln z)^{\theta - 1} / z$
\\[3ex]
Inverse Generator Functions:
\\[1ex]
$\psi_\theta^{-1}(z)$ &  $ \displaystyle (\theta z + 1)^{-1/{\theta}}$     &   $ \displaystyle   - \ln\{  e^{-z}  ( e^{-\theta} - 1) + 1\} / \theta $ & $\exp(- z^{ 1/ \theta})$
\\[2ex]
$(\psi_\theta^{-1})' (z)$  &$ \displaystyle  - (\theta z + 1)^{-1/{\theta} - 1} $ & $  \displaystyle  \frac{ e^{-z}  (e^{-\theta} - 1)}  { \theta \{ e^{-z}   (e^{-\theta} - 1) + 1\} }$ & $  \displaystyle  -  z^{1/\theta - 1} \exp(- z^{1/\theta}) / \theta $ 
\\[2ex]
$(\psi_\theta^{-1}) '' (z)$ &$ \displaystyle  (\theta + 1)\{(\theta z + 1)^{-1/{\theta} - 2} \}$ &
$ \displaystyle  -   \frac{e^{-z}  (e^{-\theta} - 1)}{\theta \{e^{-z} (e^{-\theta}  - 1) + 1\}^2}$ & $\displaystyle   z^{1/{\theta} - 2} \exp(-z^{1/\theta})(z^{ 1/{\theta} } - 1 + \theta)     / \theta^2$
\\[1ex]

\hline
\end{tabular}
}
\end{center}
\end{table}


\subsection{Estimation of the conditional marginal survival functions}	
\label{estimationS}

Since the terminal event time is independent of the censoring time, $S_{2\mid X}(\cdot \mid x)$ can be estimated using Beran's estimator \citep{beran1981nonparametric} 
\begin{equation}
\label{Beran}
\hat{S}_{2\mid X}(t|x; h) = \prod_{T_{2i} \le t, \Delta_{2i} = 1}^{}\left(1 - \frac{w_{ni}(x, h)}{1 - \sum_{j = 1}^{i - 1}w_{ni}(x, h)}\right),
\end{equation}
with the weights $w_{ni}(x, h) = K_h(x_i - x)/\sum_{j = 1}^{n}K_h(x_j - x),$ where $K_h(\cdot) = K(\cdot / h)/h$, with $K$ the kernel function and $h$ the bandwidth parameter. 

When the weights are chosen to be $w_{ni}(x, h) = 1/n$, Beran's estimator reduces to the Kaplan--Meier estimator. 
In practice, the choice of bandwidth is important to achieve reliable estimates. 
A too small (or too large) bandwidth value yields a smaller bias (variance) but a higher variance (bias). 
Here, we employ the leave-one-out cross-validated bandwidth selector in \cite{geerdens2017conditional} for Beran's estimator, along with the Epanechnikov kernel $K(z)= 3/4 (1-z^{2})_{+}$, where the subscript ``$+$'' denotes the positive part. 
This bandwidth selector minimizes
\begin{equation}
\label{eq:Beran_band2}
\sum_{i=1}^n \sum_{j=1}^n  \Delta_{2,ij} \left( \mathbbm{1}(T_{2i} \leq T_{2j}) - \hat{F}_{2|X}^{(-i)}(T_{2j}| x_i; h) \right)^2
\end{equation}
over a predefined set of bandwidth values,
where $\hat{F}_{2|X}^{(-i)}(\cdot| x_i; h) = 1 - \hat{S}_{2|X}^{(-i)}(\cdot| x_i; h)$  is the leave-one-out cross-validated Beran's estimator of the conditional marginal distribution
at bandwidth value $h$ obtained excluding the $i^{\text{th}}$ observation, and $\Delta_{2,ij}$ is the indicator of  so-called useful pair of observed times defined similarly as in \cite{geerdens2017conditional}.
Though, one can employ other kernel functions or bandwidth selectors \citep{Demin.Chimitova:2014}.

The estimation of $S_{1\mid X}(\cdot \mid x)$, however, is more involved due to dependent censoring, and requires the conditional copula $\mathcal{C}_X \{\cdot, \cdot \mid \theta(x)\}$. 
Suppose, for now, that the latter is provided.
Then, the conditional copula-graphic estimator of $S_{1\mid X} (\cdot \mid x)$ can be defined as
\begin{equation}
\label{CCGE}
\hat{S}_{1\mid X}(t|x; h) = \psi_{\theta(x)}^{-1} \left(  -\sum_{T_{1i}\le t, \Delta_{1i} = 1}  \left[ \psi_{\theta(x)}\{\hat{\Gamma}(T_{1i}^{-}|x; h)\} -\psi_{\theta(x)}\{\hat{\Gamma}(T_{1i}|x; h\} \right]\right),
\end{equation}
where $\hat{\Gamma}(\cdot \mid x; h)$ denotes the conditional survival function estimator of $T^{\ast} = \min\{Y_1, Y_2\}$ given $x$, 
obtained using Beran's estimator with the observations $\{ (T_{1i}, $ $ \Delta_{3i}); i=1,\ldots,n \}$, at the bandwidth value $h$, and $\hat{\Gamma}(t-)$ is the left-hand limit of $\hat{\Gamma}(\cdot)$ at $t$. 
The conditional copula-graphic estimator is a direct extension of the estimator in \cite{heuchenne2014likelihood} to the conditional setting, where we incorporate the covariate effect on dependence explicitly via $\theta(x)$, and replace the Kaplan--Meier estimator with Beran's estimator. The derivation of the estimator follows from \cite{rivest2001martingale}.

Since $\hat{S}_{1\mid X}(t|x; h)$ involves Beran's estimator, one faces the issue of bandwidth selection also in the conditional copula-graphic estimator. 
Here, instead of selecting a bandwidth parameter for $\hat{\Gamma}$, we adapt the cross-validated bandwidth selector in \eqref{eq:Beran_band2}  to $\hat{S}_{1\mid X}(t|x; h)$ and choose the bandwidth value that minimizes
\begin{equation}
\label{eq:Beran_band1}
\sum_{i=1}^n \sum_{j=1}^n  \Delta_{1,ij} \left( \mathbbm{1}(T_{1i} \leq T_{1j}) - \hat{F}_{1|X}^{(-i)}(T_{1j}|x_i; h) \right)^2,
\end{equation}
where $\hat{F}_{1|{X}}^{(-i)}(\cdot| x_i ; h) = 1- \hat{S}_{1|X}^{(-i)}(\cdot| x_i; h)$ is obtained using the leave-one-out cross-validated conditional copula graphic estimator. Denote by $h_1$ and $h_2$ the selected bandwidths for the estimation of $S_{1\mid X}(\cdot \mid x)$ and $S_{2 \mid X}(\cdot \mid x)$.

\subsection{Estimation of the conditional copula parameter}	

The conditional copula-graphic estimator in \eqref{CCGE} depends on the conditional copula $\mathcal{C}_X \{ \cdot, \cdot \mid \theta(x) \}$, which needs to be estimated in practice. Given the estimates  $\hat{U}_{1i} \equiv \hat{S}_{1\mid X}(T_{1i}| x_i; h_1)$ and $\hat{U}_{2i} \equiv \hat{S}_{2\mid X}(T_{2i}| x_i; h_2)$, for $i=1,\ldots,n$ of the conditional marginal survival functions at the selected bandwidths $h_1$ and $h_2$, the copula parameter $\theta(x)$ at a fixed point $x$ can be estimated using the local likelihood estimation. This approach has been previously employed in \cite{Acar.Craiu.Yao:2011} for complete data, and in \cite{geerdens2017conditional} for bivariate right-censored data.

Since the range of the copula parameter, $\Theta$ is restricted for most Archimedean copula families, the local likelihood inference is typically built using the 
re-parametrization $\theta(x) = g^{-1}\{\eta(x)\}$, where $\eta(\cdot)$ is called the \emph{calibration function} and $g^{-1}:  \mathbb{R} \rightarrow \Theta $ is a pre-specified inverse-link function. 
Table~\ref{tab1} gives the inverse-link functions used for the Archimedean copulas considered in this paper.

Consider a covariate value $x_i$ in a neighborhood of the point $x$. 
Provided that the calibration function $\eta(\cdot)$ has the second derivative at $x$, we can approximate $\eta(x_i)$ linearly using a first-order Taylor expansion around $x$, and write $\theta(x_i) \approx g^{-1}\{\gamma_{0x} + \gamma_{1x} (x_i - x)\}$, where $\eta^{r}(x)$ denotes the $r^{\text{th}}$ derivative of $\eta$ evaluated at $x$ and $\gamma_{r,x}=\eta^{(r)}(x) / r!$ for $r=0,1$. The local pseudo copula log-likelihood function of $(\gamma_{0x}, \gamma_{1x})$ at the point $x$ is then defined as
\begin{equation}
\label{CopLocLik}	
\sum_{i=1}^n \log \left( \mathcal{L} \left[  g^{-1}\{\gamma_{0x} + \gamma_{1x} (x_i - x)\}  ; \hat{U}_{1i}, \hat{U}_{2i} \mid  \Delta_{1i}, \Delta_{2i}\right] \right) \;   K_{h_\mathcal{C}}(x_i - x),
\end{equation}
where  
$K_{h_\mathcal{C}} (\cdot)= K(\cdot /h_\mathcal{C}) /h_\mathcal{C} $, with $K$ the kernel function and $h_\mathcal{C}$ the bandwidth parameter , is used to weigh the log-likelihood contributions of the observations based on the proximity of their covariate values to $x$. 
The bandwidth parameter controls the degree of smoothing and determines the width of the local neighbourhood around the target point $x$, e.g., smaller values of $h_C$ give greater weight to observations with covariate values very close to $x$, leading to a more locally adaptive estimate, whereas larger values of $h_C$ include observations farther from $x$, resulting in smoother estimates.

The likelihood contribution of the $i^{\rm{th}}$ observation $(T_{1i}, T_{2i}, \Delta_{1i}, \Delta_{2i}, x_i)$ is defined, in terms of the Archimedean generator as a function of $v$, as
\begin{multline}
\label{lik}
\nonumber
\mathcal{L}(\theta; \hat{U}_{1i}, \hat{U}_{2i}\mid  \Delta_{1i}, \Delta_{2i} ) = \{ \psi'_\theta(\hat{U}_{1i})\}^{\Delta_{1i}} \;  \{\psi'_\theta(\hat{U}_{2i})\}^{\Delta_{2i}} \;
(\psi_\theta^{-1})^{(\Delta_{1i}+\Delta_{2i})} \\ 
\times \left\{\psi_\theta(\hat{U}_{1i} )+ \psi_\theta(\hat{U}_{2i}) \right\},
\end{multline}
where $\psi'_\theta(z)$ is the first derivative of the Archimedean generator and $(\psi_\theta^{-1})^{(k)}(z)$ denotes the $k^{\rm{th}}$ derivative of the inverse of the Archimedean generator. Maximizing \eqref{CopLocLik} with respect to $(\gamma_{0x}, \gamma_{1x})$ yields the local linear estimates $\hat{\gamma}_{0x} = \hat{\eta}(x)$ and $\hat{\gamma}_{1x} = \hat{\eta'}(x)$, from which one gets $\hat{\theta}(x) = g^{-1}\{ \hat{\eta}(x)\}$.

Two aspects are noteworthy regarding this estimation procedure. 
First, as in Beran's estimator, the estimation performance depends on the value of the bandwidth parameter $h_\mathcal{C}$. 
Following \cite{Acar.Craiu.Yao:2011}, we use the leave-one-out cross-validated log-likelihood criterion to select $h_\mathcal{C}$ in our implementations. 
Second, although seldom, the likelihood contributions in \eqref{CopLocLik} can be infinite or undefined when $(\hat{U}_{1i} , \hat{U}_{2i}) \in \{0,1\}$. 
We exclude these observations when performing the local likelihood estimation. 
\cite{heuchenne2014likelihood} also employed this strategy when estimating the constant copula parameter in their unconditional copula-graphic estimator.

In fact, if one is willing to assume that the conditional copula parameter does not change with the covariate value, i.e., $\theta(x_i) =\theta$, the estimate $\hat{\theta}$ can be obtained by maximizing the pseudo copula log-likelihood 
function
\begin{equation}
\label{CopLik}	
\sum_{i=1}^n \log \left\{ \mathcal{L} \left( \theta ; \hat{U}_{1i}, \hat{U}_{2i} \mid  \Delta_{1i}, \Delta_{2i}\right) \right\}.
\end{equation}
In our proposal, we refer to this approach as the \emph{simplified} conditional copula graphic estimator, and denote it by CCGE$1$. 
The general estimator accounting for the covariate effect on the dependence structure is denoted by CCGE$2$. This distinction may be viewed analogously to that between GEE1 and GEE2 in the generalized estimating equations framework, where GEE2 explicitly models the dependence structure, whereas GEE1 treats dependence as a nuisance.

\subsection{Iterative Estimation Algorithm}

Since the estimation of the conditional marginal survival function of the non-terminal event requires the conditional copula, 
and the fitting of the conditional copula requires the estimates of the conditional marginal survival functions, we define a sequential iterative estimation algorithm, 
in which the updates of the conditional copula parameter function $\theta(\cdot)$ and conditional marginal survival function $S_{1\mid X}$ are alternated until convergence. 
The estimation steps are detailed in Algorithm~\ref{Algorithm1} for the general case of CCGE2.

\begin{algorithm}[h!]
{
\vspace{0.1cm}
\begin{algorithmic}

\medskip
\State {}
Given the observations $\{(T_{1i}, T_{2i}, \Delta_{1i}, \Delta_{2i}, x_i);  i = 1, \ldots, n\}$ and the initial estimate $\hat \theta^{(0)}(\cdot)$, perform the following steps:

\begin{algorithmic}[1]
\smallskip
\State{{\bf Estimation of ${S}_{2\mid X}$:}}

\noindent {\small a:}~ 
Select the bandwidth $h_2$ for Beran's estimator of ${S}_{2\mid X}$.

\noindent {\small b:}~ 
Obtain the estimates $\hat{S}_{2\mid X} (T_{2i} |x_i; h_2) \equiv \hat{U}_{2i}$ for $i = 1,\ldots,n$.

\end{algorithmic}

\medskip

\noindent  ~For $m = 1, 2, \ldots$

\smallskip

\begin{algorithmic}[1]
\makeatletter
\setcounter{ALG@line}{1}
\makeatother
\smallskip

\State{ {\bf Estimation of ${S}_{1\mid X}$ given $\mathcal{C}_X (\cdot, \cdot \mid  \hat\theta^{(m-1)}(x_i))$:}}

\noindent {\small a:}~  
Select the bandwidth $h_1^{(m)}$ for the conditional copula-graphic estimator of  ${S}_{1\mid X}$. 

\noindent {\small b:}~  
Obtain the estimates $\hat{S}^{(m)}_{1\mid X} (T_{1i} | x_i; h_1^{(m)}) \equiv \hat{U}^{(m)}_{1i}$ for $i = 1,\ldots,n$.

\end{algorithmic}

\medskip

\begin{algorithmic}[1]
\makeatletter
\setcounter{ALG@line}{2}
\makeatother
\smallskip

\State{\bf Estimation of $\theta(\cdot)$ given $\{ ( \hat{U}^{(m)}_{1i}, \hat{U}_{2i}, x_i); i=1, \ldots, n \}$:}

\noindent {\small a:}~   
Select the bandwidth $h_C^{(m)}$ for  estimating the conditional copula parameter.

\noindent {\small b:}~ 
Obtain the estimates $\hat{\theta}^{(m)}(x_i)$ for $i=1,\ldots,n$ using the local likelihood estimation.

\end{algorithmic}

\begin{algorithmic}[1]
\makeatletter
\setcounter{ALG@line}{3}
\makeatother
\smallskip
\State Repeat steps $2-3$, until convergence in $\hat {S}_{1\mid X}$ is achieved.     
\end{algorithmic}

\end{algorithmic}

\caption{Iterative estimation algorithm for the conditional copula-graphic estimator.}
\label{Algorithm1}
}

\end{algorithm}

The data-driven bandwidth selectors can be computationally demanding.  To reduce computational cost, one can opt out bandwidth selection in iterations of Step~2a and Step~3a. 
In our implementations, we performed bandwidth selection in these steps until two consecutive iterations return the same bandwidth value. Convergence is typically achieved in 3-4 iterations. 

The simplified conditional copula-graphic estimator (CCGE1) is obtained under the conditional copula model with a constant parameter in a similar fashion. 
This estimator employs maximum pseudo-likelihood estimation in Step~3b, and does not require bandwidth selection in Step~3a. 

Since the estimator is obtained through an iterative procedure and further includes nonparametric estimation of some model components, establishing its asymptotic properties, including consistency, is challenging.  
Therefore, in the next section, we investigate its finite-sample convergence properties through a simulation study.

\section{Simulation Study}
\label{s:simulation}

We evaluate the performances of the proposed conditional copula-graphic estimator (CCGE2) and its simplified version (CCGE1)  in comparison to the unconditional copula-graphic estimator (CGE) in a simulation study.
We generate the covariate values $x_i$ from $\rm{Uniform} (0, 1)$ and specify the conditional marginal survival functions using the exponential model
$$
S_{j \mid X}(t_{ji}) = \exp(-\lambda_j  t_{ji} \: \exp(\beta_j x_i)),
$$
where $\lambda_j$ is a constant and $\beta_j$ is the coefficient of the covariate, for $j=1,2$. 
For the specification of the covariate effects on the marginal survival functions and the dependence, we consider the following data generating models (DGM):
\begin{itemize}
\item[] DGM~1:~~ $\beta_1 = \beta_2 = 0$ ~and~ $\tau\in \{0.2, 0.5, 0.8\}$
\item[] DGM~2:~~ $\beta_1 = \beta_2 = 1$ ~and~ $\tau\in \{0.2, 0.5, 0.8\}$
\item[] DGM~3:~~ $\beta_1 = \beta_2 = 1$ ~and~ $\tau (x) = 2(x-0.5)^2+ 0.3$ with $\tau \in (0.30, 0.80)$.
\end{itemize}
DGM~1 describes the situation where the covariate has no effect on the marginal survival functions or dependence. 
In DGM~2,  the covariate affects only the marginal survival functions; and in DGM~3, it affects both the marginal survival functions and dependence.
The models were specified using Kendall's tau, with the copula parameter $\theta(\cdot)$ obtained via the conversions in Table~\ref{tab1}.

Under each model, we generated data $\{(U_{1i}, U_{2i} \mid x_i): i = 1, 2, \ldots, n\}$ of size $n = 100$ and $200$ from the Clayton, Frank and Gumbel families with the corresponding copula parameter $\theta(x)$. 
We then obtained the event times $Y_{1i} = S_{1\mid X}^{-1}(U_{1i})$ and  $Y_{2i} = S_{2\mid X}^{-1}(U_{2i})$ from the copula data  using the inverse-cdf method under the exponential model.
We set $\lambda_1 =1$ and determined $\lambda_2$ value so that the non-terminal event has no (approximately 0\%), a low (approximately 25\%) and a moderate (approximately 50\%) censoring rate. 
The censoring variable $Z$ was generated from Uniform$(0, b)$ distribution, where $b$ was chosen so that $P(Y_2>Z) = 0.20$ under each setting. 
The observed data $\{(T_{1i}, T_{2i}, \Delta_{1i}, \Delta_{2i}, x_i), i = 1, \ldots, n\}$ were obtained as discussed in Section~\ref{s:model}.
We repeated the experiment $M = 1000$ times under each setting.

For each generated sample, we fitted CGE, CCGE1, and CCGE2. The estimation performance is evaluated using the integrated squared bias and the integrated mean square error, defined as
$$
\text{ISB}(\hat{\alpha}) = \int_{X}^{} [\E[\hat{\alpha}(x)] - \alpha(x)]^2 dx \qquad \text{and} \qquad
\text{IMSE}(\hat{\alpha}) = \int_{X}^{}{\E[\{\hat{\alpha}(x) - {\alpha}(x)\}^2]}dx, 
$$
where $\hat{\alpha}(x)$ stands for either $\hat{\tau}(x)$ or $\hat{S}_{1\mid X}(\cdot\mid x)$. In these evaluations, we consider a sequence of $x$ values equally spaced between $0$ and $1$ with a step size of $0.05$.

For brevity, we report the estimation results under the Frank copula with $\tau=0.5$ for DGM~1 and DGM~2, and with $\tau = \tau(x)$ for DGM~3 for the case with sample size $n = 100$. 
The results under the Clayton and Gumbel copulas and  for sample size $n=200$ can be found in the Supplementary Material available online.

\begin{table}
	\centering

	\caption{ Mean, Integrated Squared Bias ($\text{IBias}^2$) and Integrated Mean Square Error (IMSE) (multiplied by 100) of the $\hat{S}_{1 \mid X}(.)$ at different quantiles over $1000$ Monte Carlo samples under Frank copula with sample size $n = 100$.}
	\label{tab2}
	\vspace{0.2cm}
	\scalebox{0.77}{
			\begin{tabular}{@{\hspace{-2pt}}c@{\hspace{8pt}}c@{\hspace{5pt}}c@{\hspace{2pt}}c@{\hspace{5pt}}c@{\hspace{5pt}}c@{\hspace{8pt}}c@{\hspace{5pt}}c@{\hspace{5pt}}c@{\hspace{8pt}}c@{\hspace{5pt}}c@{\hspace{5pt}}c}
			\hline\\ [-1.5ex]
			&\multirow{2}{*}{p}	&Censoring  &\multicolumn{3}{c}{CGE}& \multicolumn{3}{c}{CCGE$1$}& \multicolumn{3}{c}{CCGE$2$}\\
			&& \multicolumn{1}{c}{Rate}&$E(\hat{S}_{1}(\cdot))$& $\text{IBias}^2$& IMSE&$E(\hat{S}_{1\mid X}(\cdot))$& $\text{IBias}^2$& IMSE &$E(\hat{S}_{1\mid X}(\cdot))$& $\text{IBias}^2$&IMSE\\
			\hline\\[-1.5ex]
			\multirow{10}{*}{DGM~1}&	\multirow{3}{*}{0.1}&$0\%$&0.100&0.000&0.086&0.100&0.000&0.121&0.100&0.000&0.121\\
			&&$25\%$&0.101&0.000&0.112&0.101&0.000&0.160&0.101&0.000&0.160\\
			&&$50\%$&0.098&0.000&0.298&0.098&0.000&0.367&0.100&0.001&0.395\\[1ex]
			&\multirow{3}{*}{0.5}&$0\%$&0.499&0.000&0.258&0.500&0.000&0.388&0.500&0.000&0.386\\
			&&$25\%$&0.501&0.000&0.292&0.502&0.000&0.444&0.502&0.000&0.444\\
			&&$50\%$&0.500&0.000&0.361&0.499&0.000&0.505&0.497&0.001&0.542\\[1ex]
			&\multirow{3}{*}{0.9}&$0\%$&0.900&0.000&0.089&0.900&0.000&0.122&0.900&0.000&0.122\\
			&&$25\%$&0.900&0.000&0.098&0.900&0.000&0.140&0.900&0.000&0.140\\
			&&$50\%$&0.900&0.000&0.106&0.899&0.000&0.150&0.898&0.000&0.157\\
			&&&&&&&&&&\\
			\multirow{10}{*}{DGM~2}&\multirow{3}{*}{0.1}&$0\%$&0.123&0.434&0.536&0.107&0.048&0.242&0.107&0.048&0.243\\
			&&$25\%$&0.125& 0.446&0.578&0.110&0.064&0.319&0.107&0.048&0.242\\
			&&$50\%$&0.124&0.505&0.794&0.108&0.150&0.622&0.108&0.157&0.643\\[1ex]
			&\multirow{3}{*}{0.5}&$0\%$&0.493&0.830&1.074&0.498&0.074&0.649&0.498&0.074&0.650\\
			&&$25\%$&0.494&0.825&1.096&0.499&0.090&0.715&0.498&0.073&0.652\\
			&&$50\%$&0.493&0.808&1.137&0.495&0.134&0.798&0.494&0.141&0.839\\[1ex]
			&\multirow{3}{*}{0.9}&$0\%$&0.893&0.080&0.175&0.898&0.006&0.175&0.898&0.006&0.174\\
			&&$25\%$&0.893&0.080&  0.184&0.898&0.006&0.191&0.898&0.006&0.175\\
			&&$50\%$&0.893&0.082&0.197&0.897&0.010&0.202&0.896&0.011&0.212\\
			&&&&&&&&&&\\
			\multirow{10}{*}{DGM~3}&\multirow{3}{*}{0.1}&$0\%$&0.123&0.434&0.535&0.107&0.048&0.243&0.107&0.048&0.243\\
			&&$25\%$&0.123&0.427&0.545&0.108&0.051&0.278&0.108&0.052&0.276\\
			&&$50\%$&0.126&0.496&0.807&0.106&0.110&0.617&0.105&0.102&0.619\\[1ex]
			&\multirow{3}{*}{0.5}&$0\%$&0.493&0.831&1.074&0.498&0.073&0.651&0.497&0.074&0.652\\
			&&$25\%$&0.491&0.838&1.103&0.495&0.087&0.690&0.494&0.087&0.686\\
			&&$50\%$&0.490&0.820&1.161&0.489&0.139&0.837&0.483&0.154&0.888\\[1ex]
			&\multirow{3}{*}{0.9}&$0\%$&0.893&0.080&0.175&0.898&0.006&0.176&0.898&0.006&0.175 \\
			&&$25\%$&0.894&0.080&0.183&0.898&0.007&0.184&0.898&0.007&0.184\\
	&&$50\%$&0.894&0.080&0.193&0.897&0.011&0.201&0.895&0.015&0.223\\
			\hline
	\end{tabular}}
\end{table}

Table~\ref{tab2} summarizes the estimation results for ${S}_{1\mid X}(\cdot\mid x)$ at three different quantiles $p = 0.1, 0.5$ and $0.9$, and censoring rates $0\%, 25\%$ and $50\%$. 
In our evaluations, we consider $t_p$,  the $p^{\rm{th}}$ quantile of the conditional survival function $S_{1\mid X}$, with  ${S}_{1\mid X}(t_p \mid x) = p$. Hence, $t_{0.1}$ is located in the upper tail where most of the censoring occurs, while $t_{0.9}$ is in the lower tail of the distribution.

When there is no covariate effect, all three estimators have negligible bias.
However, compared to the ideal estimator CGE under this setting, the conditional copula-graphic estimators exhibit higher variability.
This suggests that incorporating the covariate information when there is no need may lead to an efficiency loss.

On the other hand, failing to account for the covariate effect when it is present in the margins and/or dependence can yield a considerable bias in the estimation of ${S}_{1\mid X}(\cdot\mid x)$. 
Under both DGM~2 and DGM~3, we observe that CGE consistently overestimates upper tail quantiles~($p=0.1$) and underestimates the lower tail quantiles~($p=0.9$), and this holds for all censoring rates. 
The results further indicate that the bias magnitude is much higher when estimating the survival probability at middle and upper tail quantiles ($p=0.5, 0.1$), while it is negligible at lower tail quantiles ($p=0.9$). 
Both CCGE$1$ and CCGE$2$  yield comparatively small integrated square bias and integrated mean square error under these cases. 
As expected, the estimation performance of all three estimators deteriorates with increasing censoring rate of the non-terminal event.
Similar conclusions are reached under other settings (see the Supplementary Material available online).

While these conclusions are based on integrated quantities over the range of the covariate, a detailed look at the estimation results across different covariate values further reveals that under DGM~2 and DGM~3, all three estimators but especially CGE incur, on average,  a negative bias (underestimation) of the conditional survival function at small covariate values and a positive bias (overestimation) at large covariate values (see Figure~S4 in the Supplementary Material). 
On the other hand, the estimates are on target for all three estimators across all covariate values under DGM~1. 
All three estimators show higher variability of the survival probability estimates at median  compared to the estimates at tail quantiles. 
We also observe wider confidence intervals for CCGE1 and CCGE2, especially at the boundary values for the covariate due to the nonparametric nature of these estimators. 

Overall, the results suggest that failing to account for covariate effects in the margins can be detrimental when estimating the survival function of the non-terminal event. However, the same is not true for the dependence structure. 
To our surprise, CCGE$1$ is found to perform equivalently well in estimating the conditional marginal survival function of the non-terminal event time as CCGE$2$, even when the dependence structure changes with covariate (e.g., under DGM~3). 
Hence, if interest is solely in the estimation of the conditional survival function of the non-terminal event, one can conveniently use CCGE$1$, which makes a working \emph{constant copula} assumption.

On the other hand, if the effect of covariate on the dependence structure is also of interest,  we recommend using CCGE$2$. 
As can be seen in Figure~\ref{fig1}, Kendall's tau estimates obtained from CCGE$2$  coincide with the true Kendall's tau under each model, 
whereas CGE and CCGE$1$ fail to capture the non-constant (convex) dependence pattern under DGM~3. 
We observe wider confidence intervals for Kendall's tau at higher censoring rates irrespective of the data generation model. 
Similar results are obtained under the Clayton and Gumbel copulas (see the Supplementary Material available online).

\begin{figure}
\centering
\caption{Mean, $5^{\text{th}}$ and $95^{\text{th}}$ quantiles of Kendall's tau estimates under Frank copula with sample size $n = 100$. Data is generated from DGM~1(top row), DGM~2 (middle row) and DGM~3 (bottom row) with no (left column), low (middle column) and moderate (right column) censoring rates of non-terminal event. Dashed lines represent the Kendall's tau estimates from CGE (red), CCGE$1$ (green), CCGE$2$ (blue); Dotted and solid lines represents the quantiles and true Kendall's tau.}\vspace{-0.4cm}
\label{fig1}
\begin{tabular}{@{\hspace{-2pt}}c@{\hspace{-26pt}}c@{\hspace{-26pt}}c}

\vspace{-2cm}
\includegraphics[height =2.6in, width =1.9in ]{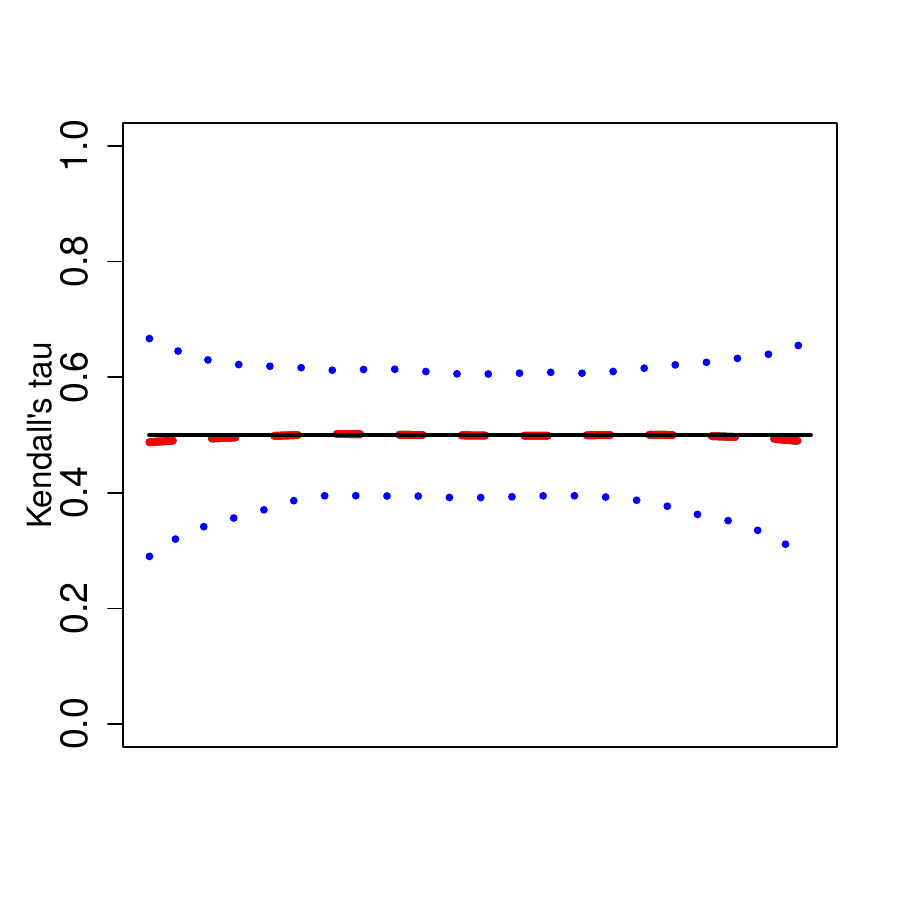} &
\includegraphics[height =2.6in, width =1.9in ]{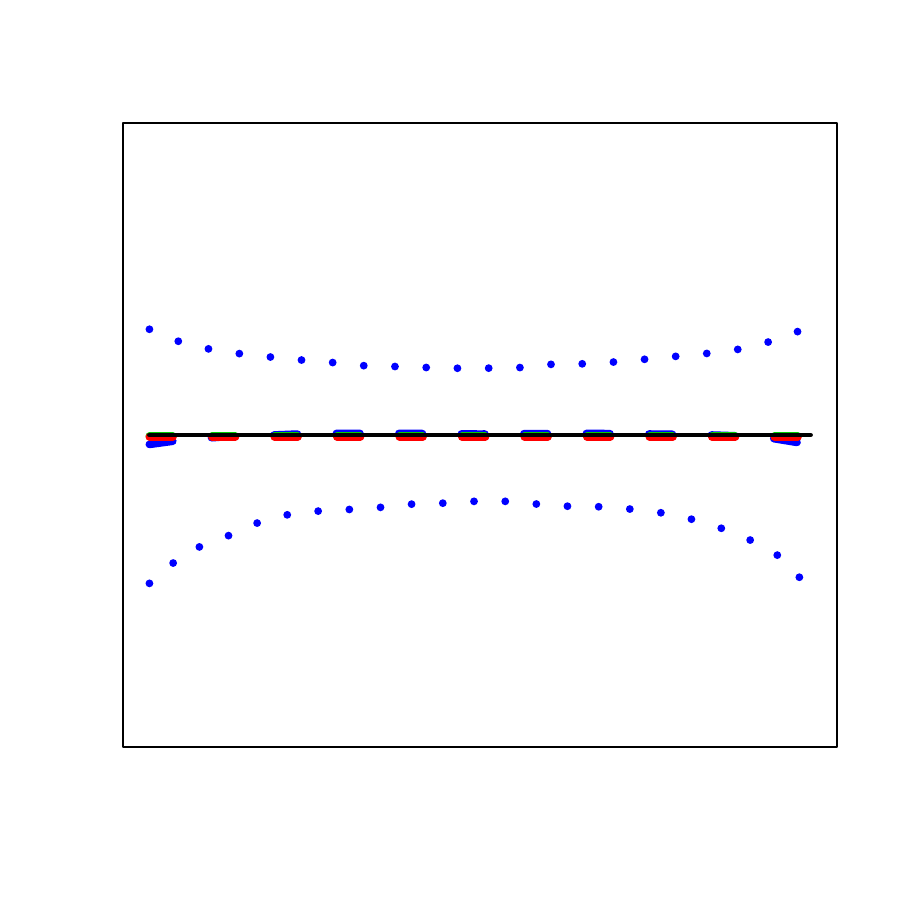} & 
\includegraphics[height =2.6in, width =1.9in ]{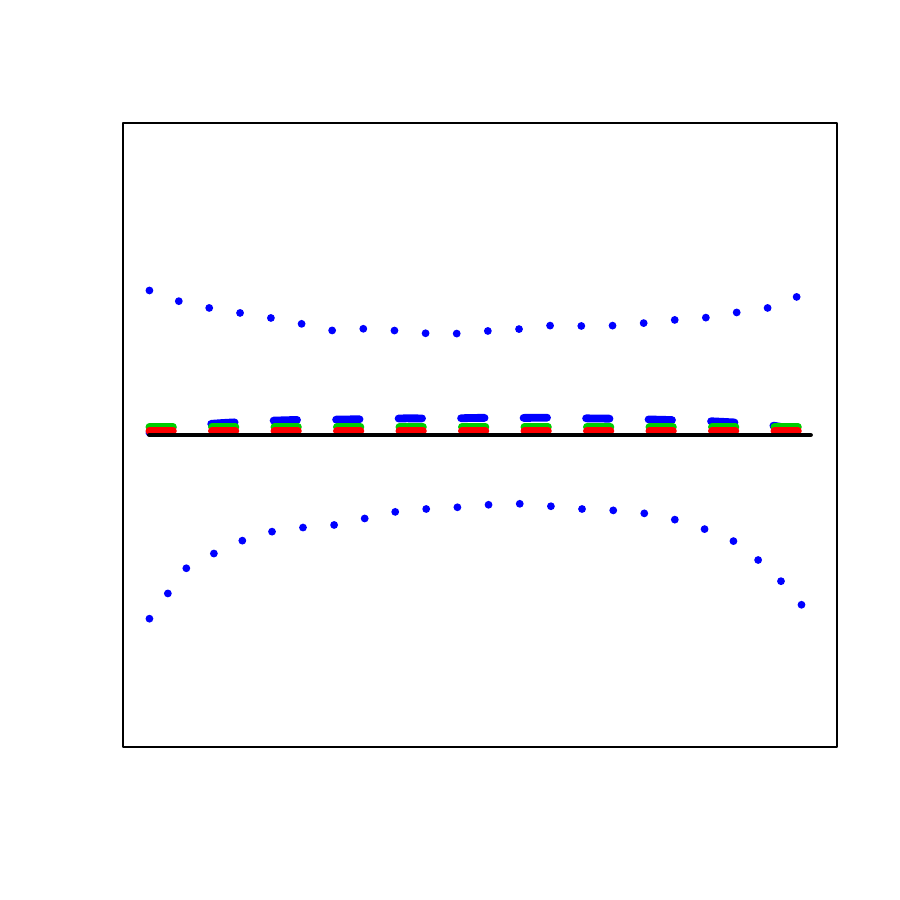}
\\	
		
\vspace{-2cm}
\includegraphics[height =2.6in, width =1.9in ]{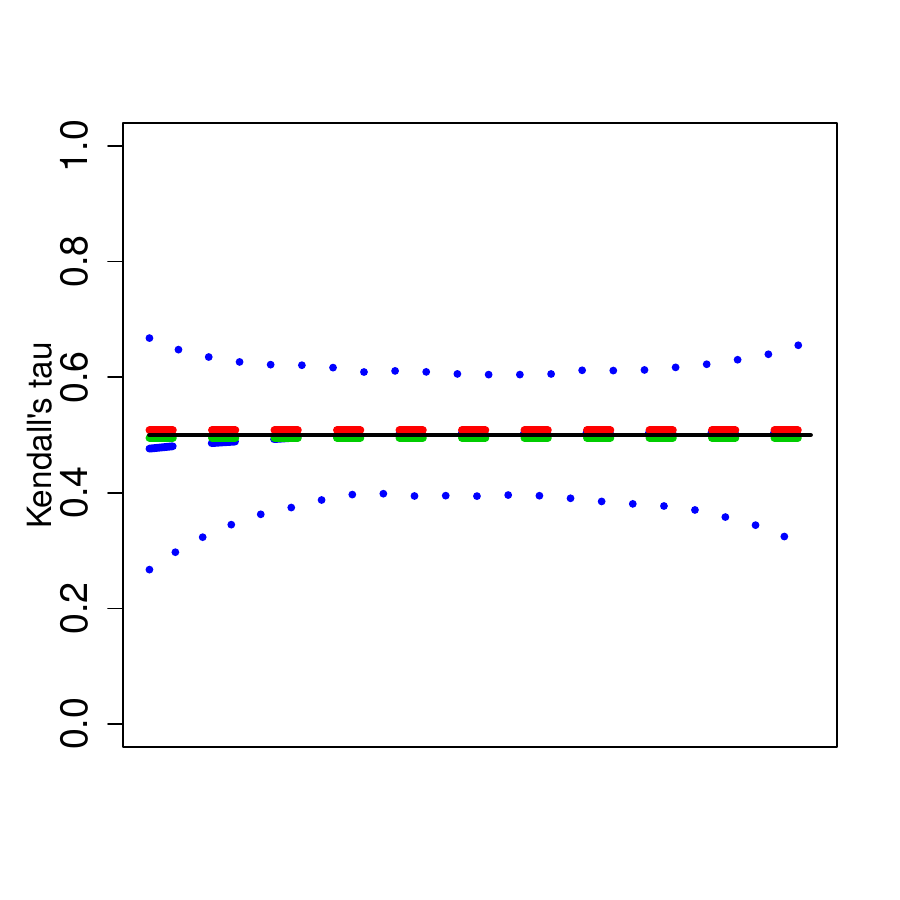} & 
\includegraphics[height =2.6in, width =1.9in ]{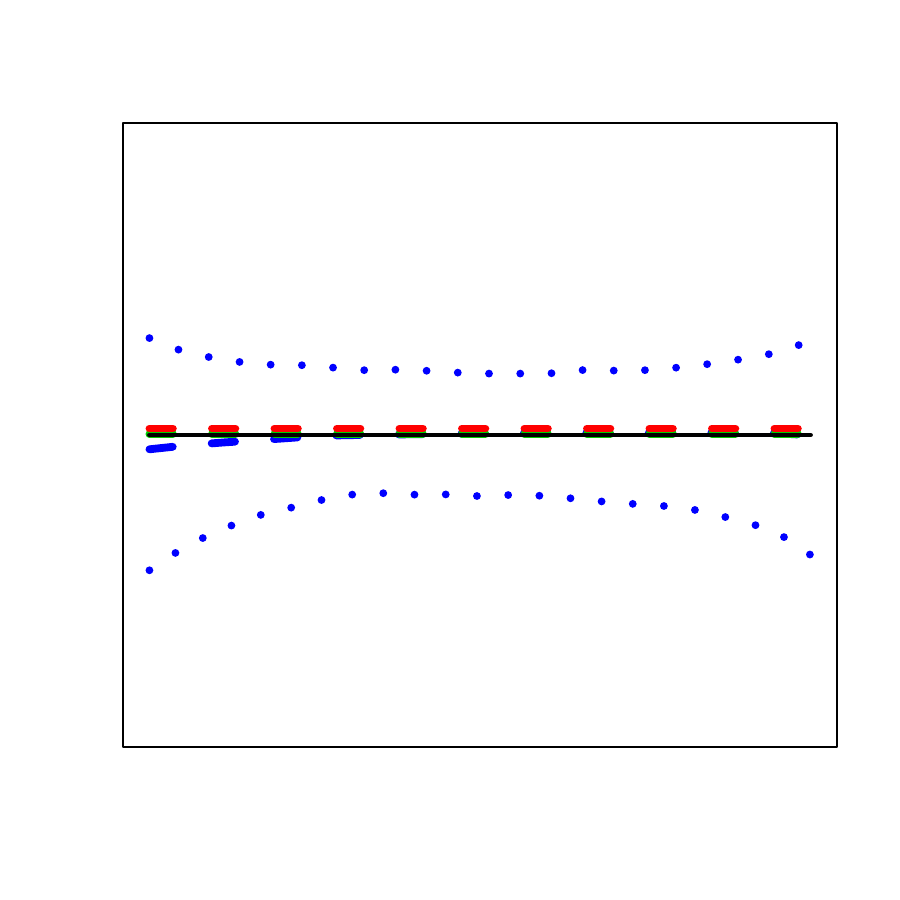} & 
\includegraphics[height =2.6in, width =1.9in ]{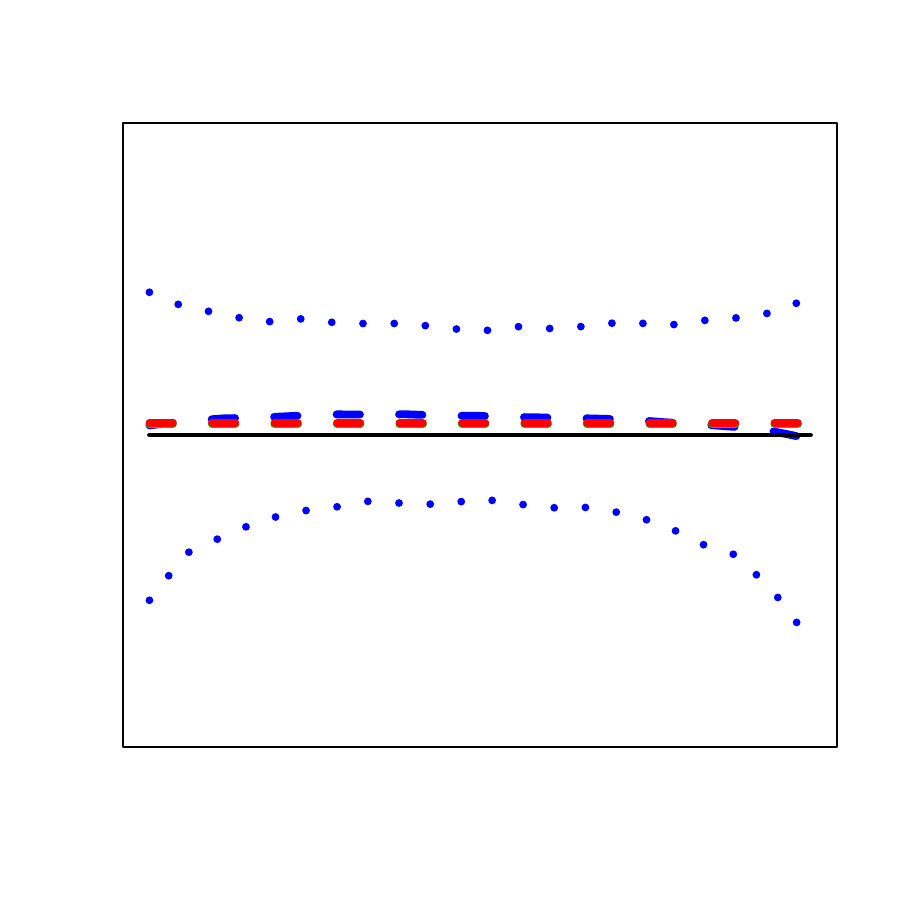}
\\
	
\includegraphics[height =2.6in, width =1.9in ]{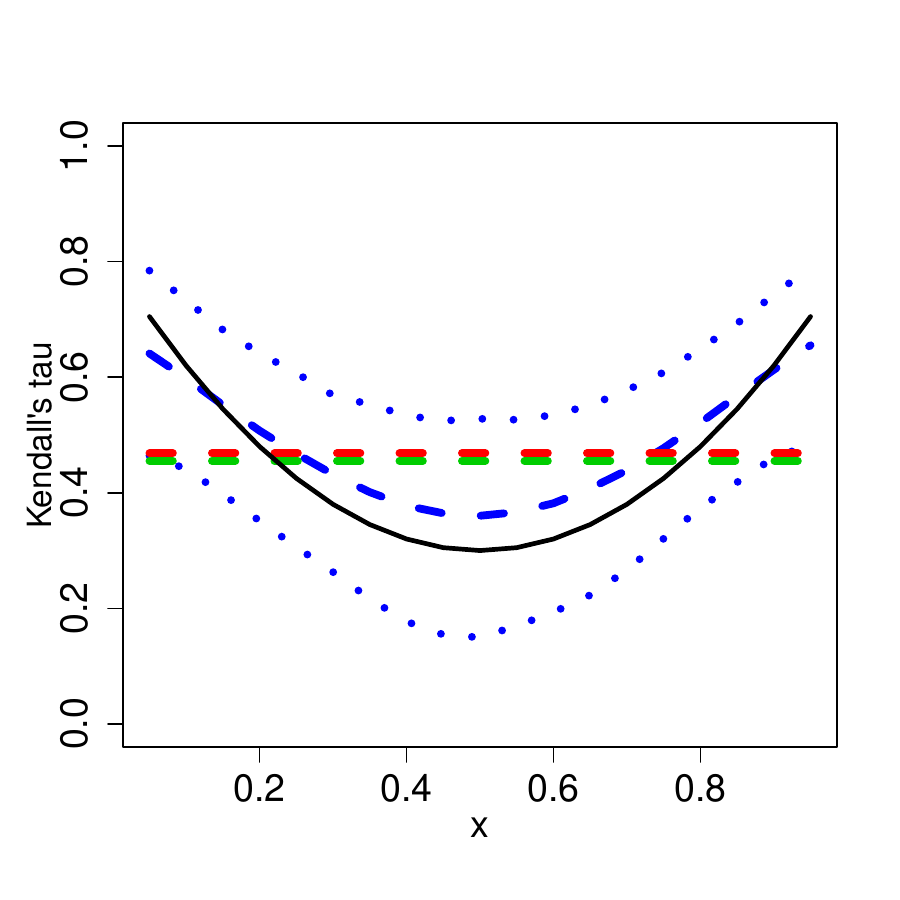}&  
\includegraphics[height =2.6in, width =1.9in ]{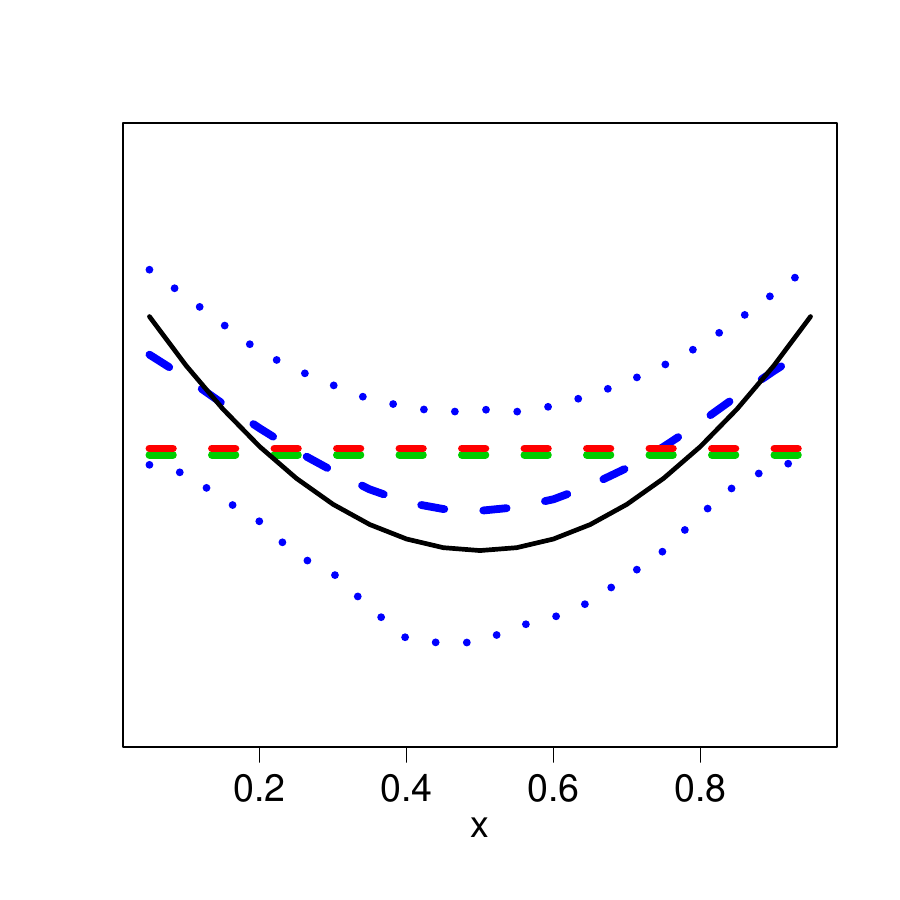} & 
\includegraphics[height =2.6in, width =1.9in ]{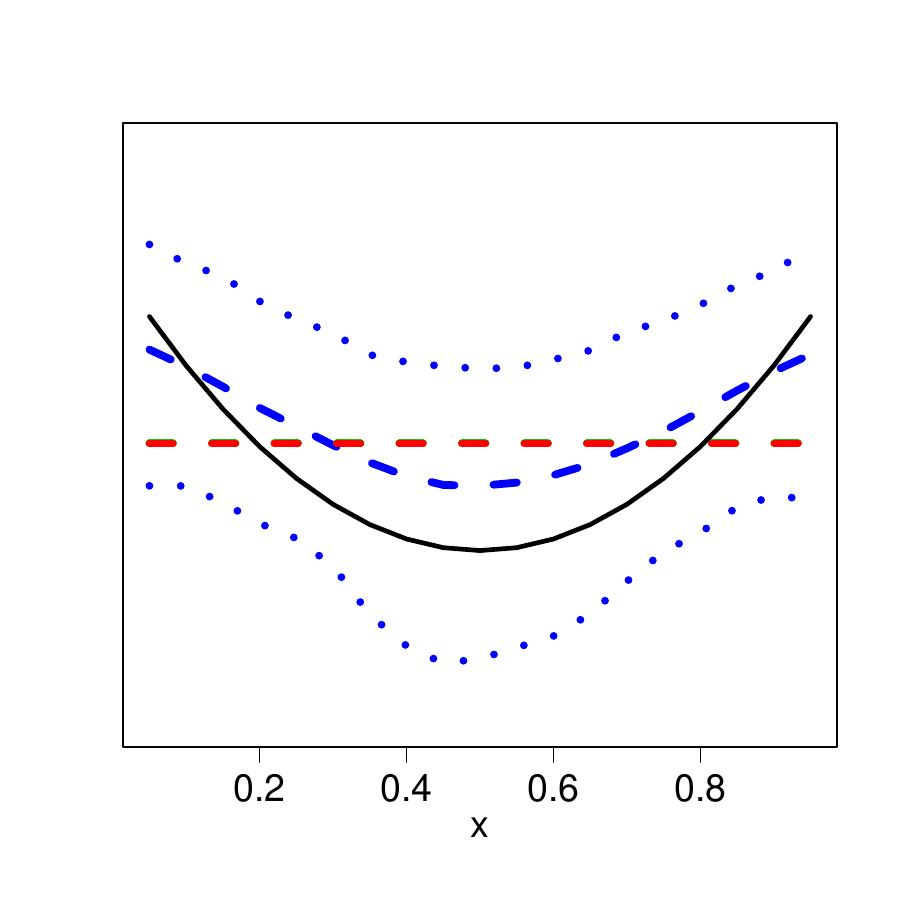}\\
\end{tabular}
\end{figure}

\section{Data Application}
\label{s:data}

In this section, we demonstrate our methods on two real data sets, which were previously analyzed in \cite{wang2003estimating} in an unconditional setting. Here, we investigate potential covariate effects in these applications using the conditional copula-graphic estimators.

\subsection{Stanford Heart Transplant Data}

We first analyze a subset of the data collected from the Stanford Heart Transplant Program \citep{clark1971cardiac}.
The analysis subset consists of $n = 103$ end-stage heart disease patients with data on the heart transplant waiting time (in days),
transplant status (1=transplant, 0=censored), time to death (in days), death status (1=dead, 0=censored) and age at acceptance into the program (in years).  
Among these patients, $69$ ($67\%$) received heart transplant but only $24$ survived during the study period. 
The remaining $34$ ($33\%$) patients did not have a heart transplant, and among them only $4$ survived. 
The age of patients ranges from $8$ to $64$ with median age $47$ years. 
More details about this program can be found in \cite{clark1971cardiac}, as well as in \cite{crowley1977covariance}, and \cite{aitkin1983reanalysis}.

Our aim is to infer whether the heart transplant waiting time changes with the age of patient, or whether the latter has any effect on the dependence between transplant waiting time and lifetime of patients. 
We employed the unconditional and conditional copula-graphic estimators using the Clayton, Gumbel and Frank copulas (see Table~\ref{tab3}). 
All three families suggest a weak dependence between the transplant waiting time and time to death, 
with Kendall's tau estimates below 0.1, regardless of the estimator employed. 
Since the log-likelihood values under the three families are very close for each estimator, we decided to use the Frank copula when examining the covariate effect. 
This family is appealing from a modeling perspective as it allows for both negative and positive dependence. 
\cite{wang2003estimating} also used the Frank copula when analyzing this data set.

\begin{table}[h!]
\centering
\caption{Log-likelihood values and Kendall's tau estimates under the Clayton, Gumbel, and Frank families using the CGE and CCGE1 estimators for the Stanford heart transplant data.}
\label{tab3}
\vspace{0.2cm}
\scalebox{0.9}{
\begin{tabular}{l cc cc cc}
\hline
 & \multicolumn{2}{c}{Clayton} &
   \multicolumn{2}{c}{Gumbel} &
   \multicolumn{2}{c}{Frank} \\
\cmidrule(lr){2-3}\cmidrule(lr){4-5}\cmidrule(lr){6-7}
 & log-likelihood & $\hat{\tau}$ & log-likelihood & $\hat{\tau}$  & log-likelihood & $\hat{\tau}$  \\ 
  \hline
CGE & -56.06 & 0.00 & -56.21 & 0.07 & -56.24 & 0.06 \\ 
CCGE1 & -52.25 & 0.00 & -52.50 & 0.09 & -52.57 & 0.07 \\ 
   \hline
\end{tabular}
}
\end{table}

\begin{figure}[h!]
	\caption{Survival function of the heart transplant waiting time and Kendall's tau estimates obtained using CGE (red), CCGE$1$ (green) and CCGE$2$ (blue) with 95\% confidence interval (dotted) for the Stanford heart transplant data.}
	\centering
	\label{fig2}
	\begin{tabular}{c@{\hspace{-15pt}}c@{\hspace{-15pt}}c}
		\vspace{-0.3cm}
	\includegraphics[scale=0.188]{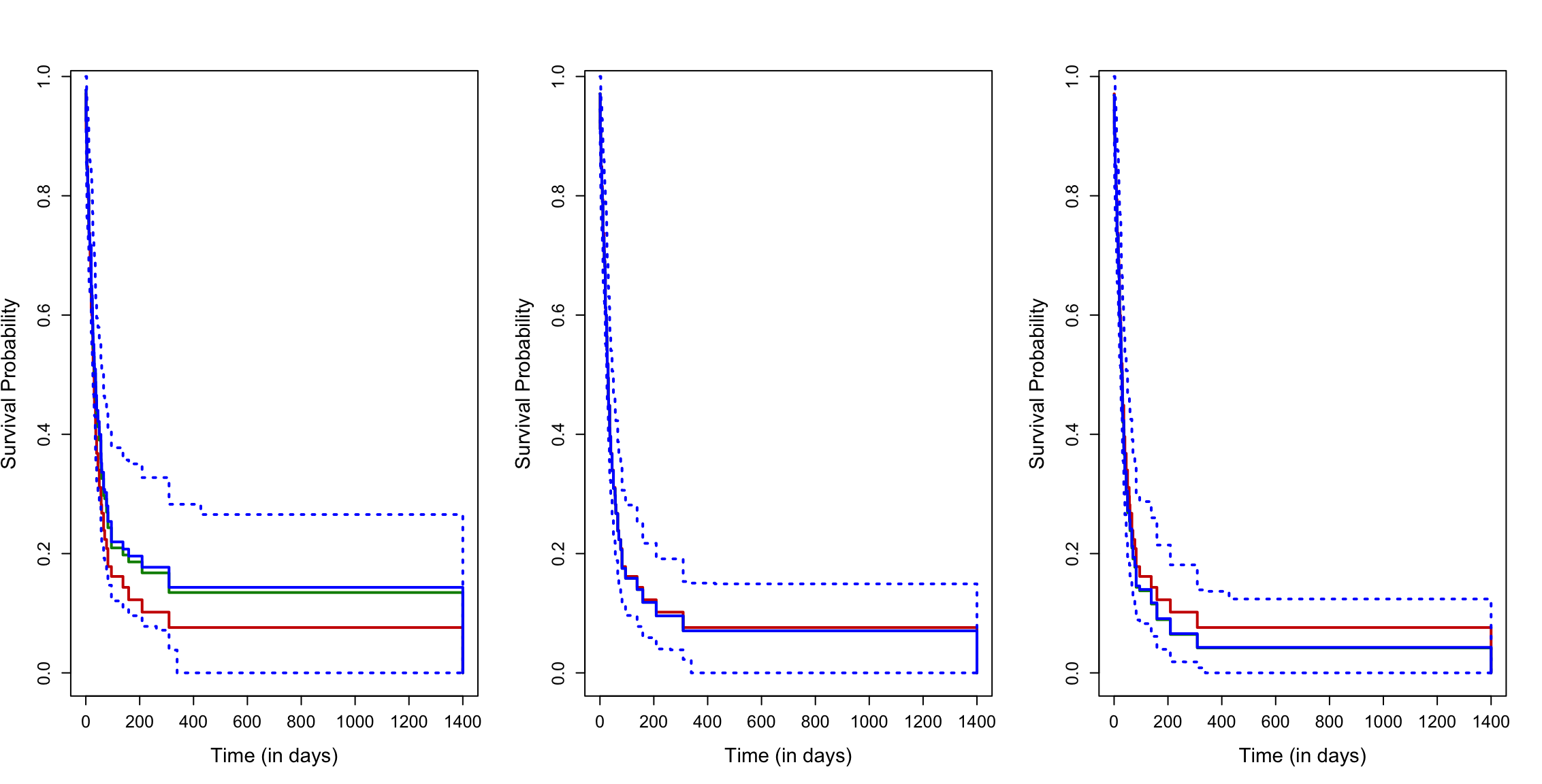} \\ 
	\\
	\end{tabular}
	\centering (a) Conditional survival function estimates for patients with age 20 (left panel), 40 (middle panel) and 60 (right panel) years at the time of acceptance into the program.
	\begin{center}
		\includegraphics[scale=0.21]{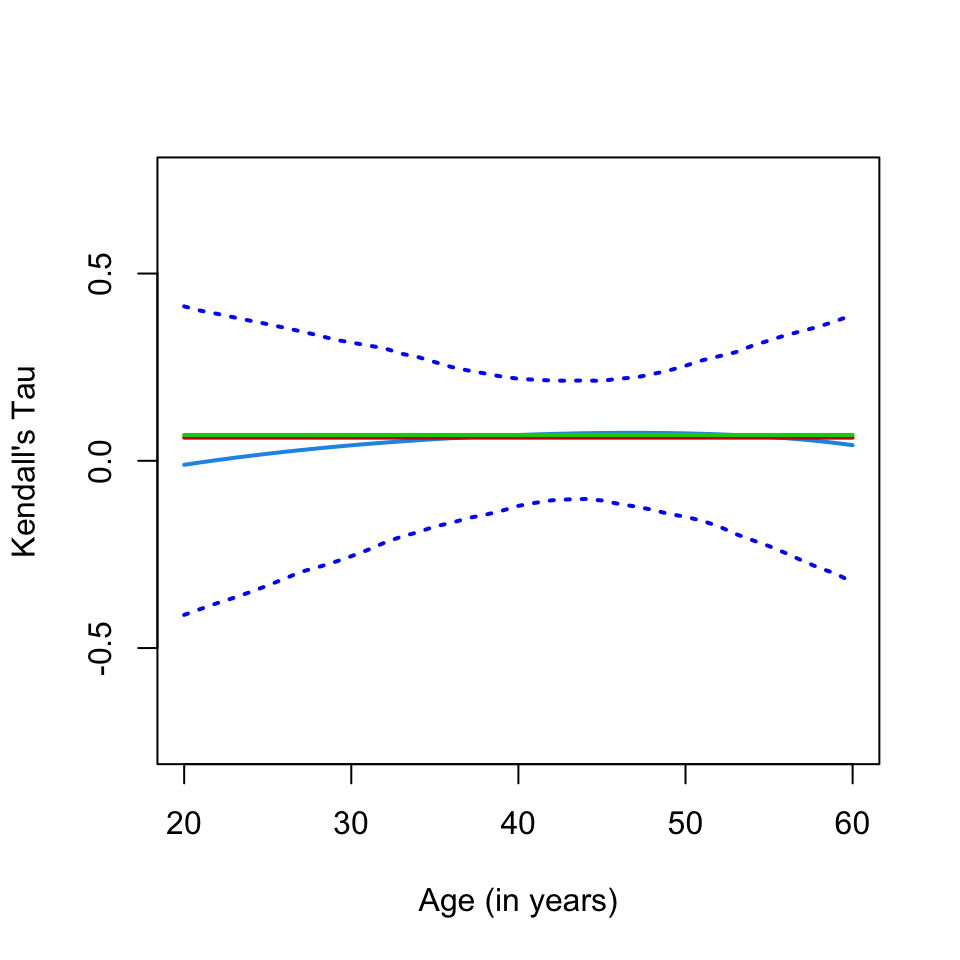}\\
		(b) Kendall's tau estimates as a function of patient's age at acceptance into the program.
	\end{center}
\end{figure}

To assess the effect of patient's age on the heart transplant waiting time, we fit the conditional copula-graphic estimators (CCGE$1$ and CCGE$2$), and compared the results with those of the unconditional copula-graphic estimator (CGE). 
The selected bandwidth values, among the pilot bandwidth values ranging from $15$ to $56$,
were $(h_1 = 37, h_2 = 27)$ for CCGE$1$ and $(h_1 = 37, h_2 = 27, h_C = 56 )$ for CCGE$2$.
The conditional survival function estimates are displayed in Figure \ref{fig2} for patients with age 20, 40 and 60 years at the time of acceptance into the program. 
To assess the uncertainty in these estimates, we use nonparametric bootstrap and construct $95\%$ bootstrap confidence intervals from the CCGE$2$ estimates obtained under the same selected bandwidth values for $B = 1000$ bootstrap samples.

For young patients the unconditional and conditional copula-graphic estimators yield very different survival probabilities, 
and ignoring patient's age results in an underestimation of the survival probability of the transplant waiting time. 
That is, young patients have a transplant waiting time longer than suggested by the unconditional analysis.
For patients at ages $40$ and $60$, all three estimators perform similarly, 
though CGE slightly overestimates the survival probability of the transplant waiting time beyond $100$ days.
The results under CCGE$1$ and CCGE$2$ were indistinguishable across the considered covariate values, supporting the conclusions  in Section~\ref{s:simulation}.

To examine the effect of patient's age on the dependence between the heart transplant waiting time and time to death, we compared Kendall's tau estimates obtained using the three estimators (see Figure \ref{fig2}). 
While CGE and CCGE$1$ yield similar constant Kendall's tau estimates ($\tau=0.06$ and $0.07$, respectively), 
CCGE$2$ indicates a slight variation in Kendall's tau across ages $8$ to $64$. 
A negative Kendall's tau is obtained for age below $25$ and above $60$. 
Since young and old people are more vulnerable in general, the longer they wait for a transplant, the higher their risk of death. 
On the other hand, for patients who are between $25$ and $60$, there is almost no association between the transplant waiting time and the survival time.

\subsection{Bone Marrow Transplant Data}

We next analyze the bone marrow transplant data \citep{klein2006survival}, available on the {\tt KMSurv} package in \textsf{R}. 
Recorded in the dataset are the time to relapse of leukaemia from transplantation (in days), 
relapse status (1=relapse, 0=censored), 
time to death from transplantation (in days), 
death status (1=dead, 0=censored) and age at bone marrow transplant (in years)  for $n = 137$ leukaemia patients who received bone marrow transplants.
Among these patients, $42$ ($31\%$) had relapse of leukaemia with only $2$ surviving during the study period. 
The remaining $94$ ($69\%$) were disease free during the study period, and among them $54$ survived.
The age of patients ranges from $7$ years to $52$ years with median age $28$ years.
Our aim is to investigate the effect of the age of patient at bone marrow transplant on their leukaemia relapse time, as well as on the dependence between time to relapse and time to death.

We employed CGE and CCGE1 under three copula families and found a strong dependence between the time to relapse and death, 
with Kendall's tau estimates of approximately 0.7 and above, regardless of the estimator used (see Table~\ref{tab4}). 
Based on the log-likelihood values, we decided to use the Frank family in further analysis, though the Clayton copula is also a reasonable choice for this dataset \citep{wang2003estimating}. 

\begin{table}[h!]
\centering

\caption{Log-likelihood values and Kendall's tau estimates under the Clayton, Gumbel, and Frank families using the CGE and CCGE1 estimators for the bone marrow transplant data.}
\label{tab4}
\vspace{0.2cm}
\scalebox{0.9}{
\begin{tabular}{l cc cc cc}
\hline
 & \multicolumn{2}{c}{Clayton} &
   \multicolumn{2}{c}{Gumbel} &
   \multicolumn{2}{c}{Frank} \\
\cmidrule(lr){2-3}\cmidrule(lr){4-5}\cmidrule(lr){6-7}
 & log-likelihood & $\hat{\tau}$ & log-likelihood & $\hat{\tau}$  & log-likelihood & $\hat{\tau}$  \\ 
  \hline
CGE & -48.40 & 0.80 & -50.99 & 0.71 & -48.03 & 0.76 \\ 
  CCGE1 & -47.04 & 0.81 & -50.59 & 0.69 & -46.70 & 0.76 \\ 
   \hline
\end{tabular}
}
\end{table}

We fit CCGE$1$ and CCGE$2$ to assess the effect of patient's age at bone marrow transplant on the leukaemia relapse time. 
The selected bandwidth parameters, among the pilot bandwidth values ranging from $5$ to $45$,
 were
$(h_1 = 45, h_2= 45)$ and $(h_1 = 45, h_2= 45, h_C = 45)$, respectively. 
The results are displayed in Figure \ref{fig3} for patients with age $20$, $30$ and $40$ years at the time of bone marrow transplantation, along with the $95\%$ bootstrap confidence intervals for CCGE$2$ obtained from $B = 1000$ bootstrap samples.

For all the three age groups, the survival function estimates of CCGE$1$ and CCGE$2$ coincide with those of CGE, indicating that patient's age at transplant has no significant effect on the leukaemia relapse time.  
However, the same is not true for the dependence between the leukaemia relapse time and the lifetime.
When we compared Kendall's tau estimates from the three estimators (see Figure \ref{fig3}), we see a visible increase in the strength of dependence with patient's age at transplant.
The relatively weaker dependence between the leukaemia relapse time and lifetime of young patients suggests that 
the life course of young patients depends on the leukaemia relapse time less than that of old patients. Note that this aspect would be missed by both CGE and CCGE$1$, which yield a strong overall dependence between the two event times, with Kendall's tau estimates, $0.761$ and $0.764$, respectively.

\begin{figure}
	\caption{Survival function of the heart transplant waiting time and Kendall's tau estimates obtained using CGE (red), CCGE$1$ (green) and CCGE$2$ (blue) with confidence interval (dotted) for the bone marrow transplant data.}
	\centering
	\label{fig3}
	\begin{tabular}{c@{\hspace{-15pt}}c@{\hspace{-15pt}}c}
		\vspace{-0.3cm}
	\includegraphics[scale=0.188]{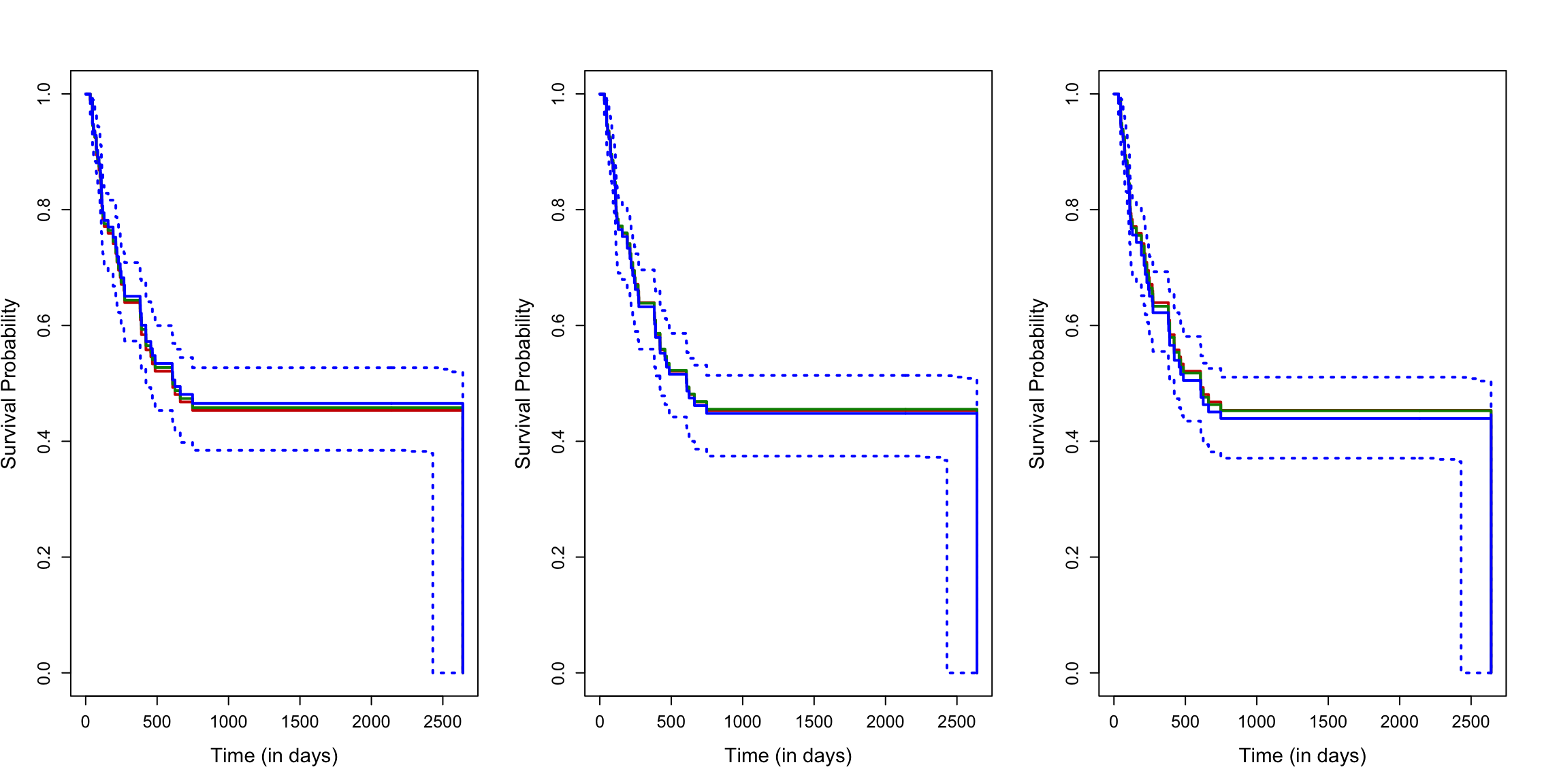} \\ \\	
	\end{tabular}
	\centering (a) Conditional survival function estimates for patients with age 20 (left panel), 30 (middle panel) and 40 (right panel) years at the time of acceptance into the program.
	\begin{center}
		\includegraphics[scale=0.21]{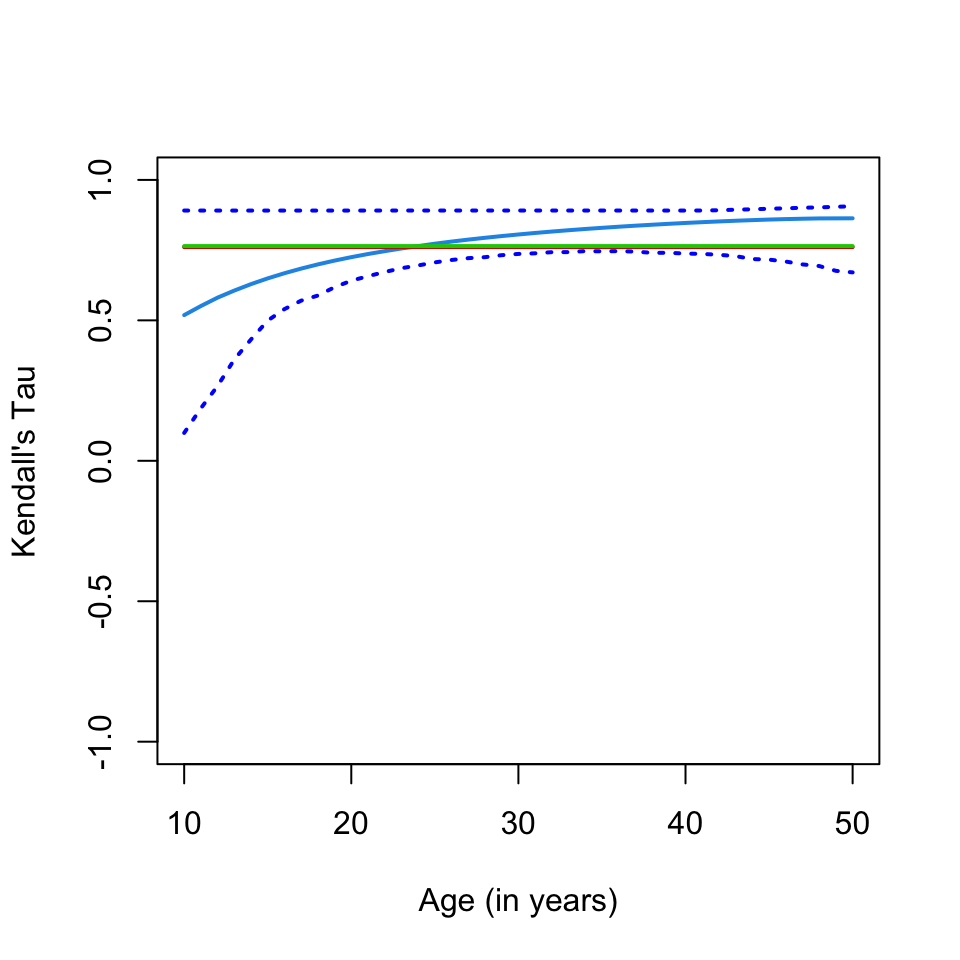}\\
		(b) Kendall's tau estimates as a function of patient's age at acceptance into the program.
	\end{center}
\end{figure}

\section{Discussion}
\label{s:conc}

In this paper, we proposed an extension of the copula-graphic estimator to allow for covariate adjustment in the analysis of semi-competing risks data. 
The proposed conditional copula graphic estimator accounts for covariate effects both  in the marginal survival functions of the non-terminal and terminal event times and in their dependence structure.
We additionally considered a simplified version of the conditional copula-graphic estimator, which makes a working \emph{constant copula} assumption, hence ignores potential covariate effects on the dependence structure.
The performance of our proposed estimators were investigated in a simulation study and compared to that of the unconditional copula-graphic estimator.

Our findings suggest that failing to account for covariate effects in the margins would yield unreliable estimates for the survival function of the non-terminal event time.
On the other hand, ignoring potential covariate effects on the dependence structure yields almost indistinguishable results for the conditional survival function of the non-terminal event time.
Hence, unless the covariate effect on the dependence structure is of interest, we recommend using the simplified conditional copula-graphic estimator.

We applied the proposed conditional copula graphic estimators to data on heart transplant and bone marrow transplant.
In the heart transplant study, we found that patient's age at acceptance into the program has an effect on the survival function of the transplant waiting time, while the dependence between waiting time and lifetime of heart disease patients appears negligible.
In the bone marrow transplant study, we did not detect any significant effect of patient's age at bone marrow transplant on survival function of the leukaemia relapse time. However, the dependence between the leukaemia relapse time and lifetime of patients shows an increase with patient's age at transplant.

The practical use of the conditional copula graphic estimators requires choosing a suitable Archimedean copula family for the conditional copula. 
While we recommend using the Frank copula as a flexible choice that allows for both negative and positive dependence, formal model selection tools need to be developed.  
In this paper, we adopted a heuristic approach to specify the copula family based on log-likelihood comparisons of the unconditional and conditional copula graphic estimators assuming a constant dependence parameter.
Our preliminary investigations (see the Supplementary Material available online) suggest that this approach correctly identifies the copula family in over 80\% of the simulated scenarios with moderate sample sizes, even when the underlying dependence parameter varies with a covariate.

The proposed conditional copula-graphic estimators can accommodate only one covariate.
This is mainly due to the additional complexity in accounting for multiple covariates in Beran's estimator. 
A future research direction is to extend the proposed estimators to settings involving multiple covariates, possibly within a penalized regression framework similar to \cite{Sun:2024}.

Although copula models are widely applied in survival analysis, copula-based predictions for censored data has received little attention. We plan to address this gap in the context of semi-competing risks, building on the ideas in \cite{Acar.Azimaee.Hoque:2019}.

\section*{Acknowledgements}
Funding in support of this work was provided by the Natural Sciences and Engineering Research Council of Canada (RGPIN 06753-2020), the Canada
Research Chairs Program, and the Canadian Statistical Sciences Institute (CANSSI) Collaborative Research Team Project to Acar.

\bibliographystyle{agsm} 
\bibliography{CCGE}

\section*{Supplemental Material}

\setcounter{table}{0}
\renewcommand{\thetable}{S\arabic{table}}%
\setcounter{figure}{0}
\renewcommand{\thefigure}{S\arabic{figure}}

This supplement contains the results of additional simulations.

\medskip
\subsection*{Simulation Results under the Frank Copula ($n = 100$)}
\label{A}
\begin{table}[H]
	\centering
	\caption{Integrated Squared Bias ($\text{IBias}^2$) and Integrated Mean Square Error (IMSE) (multiplied by 100) of the Kendall's tau estimates calculated over $1000$ Monte Carlo samples under Frank copula (No, Low and Moderate Censoring, $\tau = 0.5$ or $\tau(x)$, $n = 100$).}
	\label{tab7}
	\scalebox{0.8}{	\begin{tabular}{ccccclccccc}
			\\ [-1ex]
			\hline& \\ [-1.5ex]
			&Censoring& \multicolumn{2}{c}{CGE}&& \multicolumn{2}{c}{CCGE$1$}&& \multicolumn{2}{c}{CCGE$2$}\\
			&Rate	&$\text{IBias}^2$&  IMSE&& $\text{IBias}^2$&IMSE && $\text{IBias}^2$&IMSE\\
			\hline& \\[-1.5ex]
			\multirow{3}{*}{DGM~1}&0\%&0.002&0.259&&0.004&0.269&&0.002&0.703\\
			&25\%&0.001&0.333&&0.000&0.342&&0.004& 0.992\\
			&50\%&0.006&0.487&&0.020&0.536&&0.060&1.589\\ [1ex]
			\multirow{3}{*}{DGM~2}&0\%&0.007&0.255&&0.002&0.271&&0.006&0.714\\ 
			&25\%&0.013&0.333&&0.000&0.347&&0.007&0.734\\ 
			&50\%&0.043&0.500&&0.033&0.563&&0.075& 1.600\\    [1ex]
			\multirow{3}{*}{DGM~3}&0\%&1.819&2.124&&1.788&2.116&&0.242&3.823\\
			&25\%&1.856&2.192&&1.808&2.174&&0.288&3.964\\
			&50\%&1.910&2.494&&1.913&2.553&&0.719&4.716\\
			
			\hline
\end{tabular}}
\end{table}

\clearpage

\subsection*{Simulation Results under the Frank Copula ($n = 200$)}
\label{B}
\begin{table}[H]
	\centering
	\caption{Mean, Integrated Squared Bias ($\text{IBias}^2$) and Integrated Mean Square Error (IMSE) (multiplied by $100$) of the $\hat{S}_{1\mid X}(.)$ at different quantiles over $1000$ Monte Carlo samples under Frank copula (Low Censoring, $\tau = 0.5$ or $\tau(x)$, $n = 200$).}
	\small	
	\label{tab1}
	\scalebox{0.8}{	\begin{tabular}{p{15mm}ccccccccccccc}
			\\ [-1ex]\hline\\ [-1.5ex]
			&\multirow{2}{*}{p}	&\multicolumn{3}{c}{CGE}&& \multicolumn{3}{c}{CCGE$1$}&& \multicolumn{3}{c}{CCGE$2$}\\
			&&$E(\hat{S}_{1}(\cdot))$& $\text{IBias}^2$& IMSE&&$E(\hat{S}_{1\mid X}(\cdot))$& $\text{IBias}^2$& IMSE &&$E(\hat{S}_{1\mid X}(\cdot))$& $\text{IBias}^2$& IMSE\\
			\hline\\[-1.5ex]
			\multirow{4}{*}{DGM~1}&0.1&0.101&0.000&0.055&&0.101&0.000&0.078&& 0.101&0.000& 0.078\\
			&0.5&0.500&0.000&0.150&&0.500&0.000&0.239&& 0.500& 0.000&  0.240\\
			&0.9&0.900&0.000&0.051&&0.900&0.000&0.075&&0.900&0.000&0.075\\[1ex]
			
			\multirow{4}{*}{DGM~2}&0.1&0.124&0.434&0.498&&0.106& 0.030& 0.170&&0.106&0.030&0.170\\
			&0.5&0.492&0.833&0.970&&0.498&0.050&0.398&&0.498&0.049&0.400\\
			&0.9&0.893&0.081&0.133&&0.899&0.003&0.112&&0.898&0.003&0.113\\[1ex]
			
			\multirow{3}{*}{DGM~3}&0.1&0.121&0.414&0.471&&0.104&0.023&0.142&&0.104&0.024&0.144\\
			&0.5&0.489&0.844&0.974&&0.494&0.049&0.383&&0.495&0.051&0.383\\
			&0.9&0.893&0.081&0.131&&0.898&0.004&0.106&&0.898&0.004&0.107\\
			\hline
	\end{tabular}}
\end{table}

\begin{table}[H]
	\centering
	\caption{Integrated Squared Bias ($\text{IBias}^2$) and Integrated Mean Square Error (IMSE) (multiplied by $100$) of the Kendall's tau estimates calculated over $1000$ Monte Carlo samples under Frank copula (Low Censoring, $\tau = 0.5$ or $\tau(x)$, $n = 200$).}
	\label{tab2}
	\scalebox{0.85}{	\begin{tabular}{cccclccccc}
			\\ [-1ex]	\hline& \\ [-1.5ex]
			& \multicolumn{2}{c}{CGE}&& \multicolumn{2}{c}{CCG$1$E}&& \multicolumn{2}{c}{CCGE$2$}\\
			&$\text{IBias}^2$&IMSE&& $\text{IBias}^2$& IMSE && $\text{IBias}^2$& IMSE\\
			\hline& \\[-1.5ex]
			DGM~1&0.000&0.151&&0.000&0.151&&0.001&0.395\\
			DGM~2&0.019&0.162&&0.001&0.152&&0.004&0.401\\    
			DGM~3&1.870&2.027&&1.815&1.980&&0.146&3.743\\
			\hline
	\end{tabular}}
\end{table}

\begin{figure}[H]
	\centering
	\caption{Mean, $5^{\text{th}}$ and $95^{\text{th}}$ quantiles of Kendall's tau estimates under Frank copula (Low Censoring, $\tau = 0.5$ or $\tau(x)$, $n = 200$) from DGM~1(left), DGM~2 (middle) and DGM~3 (right). Dashed lines represent the Kendall's tau estimates from CGE (red), CCGE$1$ (green), CCGE$2$ (blue); Dotted (blue) and solid (black) lines represents the quantiles of CCGE$2$ and true Kendall's tau.}\vspace{-0.4cm}
	\label{fig1}
	\begin{tabular}{@{\hspace{-2pt}}c@{\hspace{-26pt}}c@{\hspace{-26pt}}c}
		\vspace{-0.6cm}
		\includegraphics[height=2.5in, width = 2.1in]{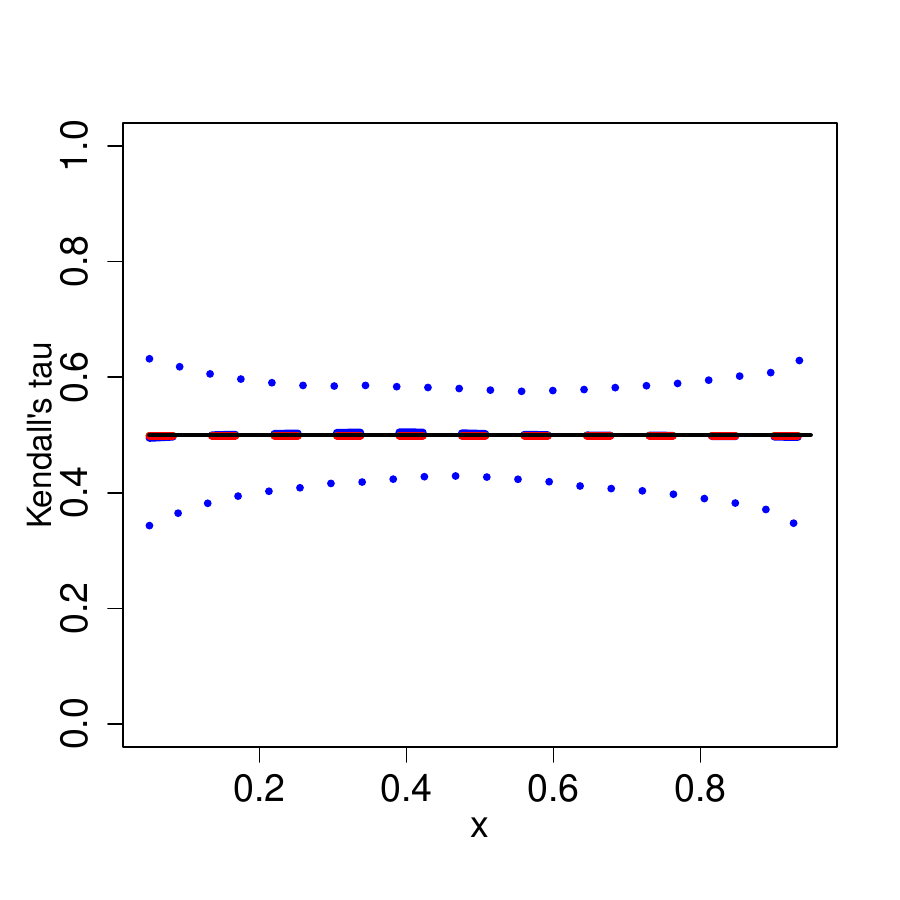} &
		\includegraphics[height=2.5in, width = 2.1in]{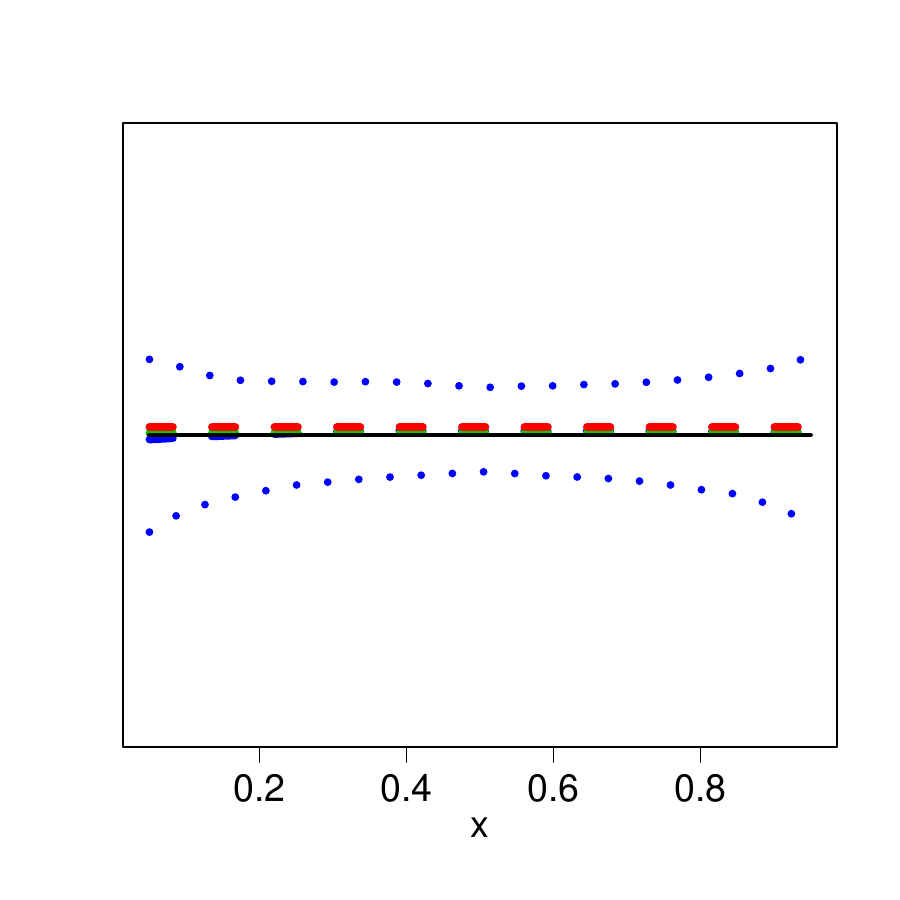} &
		\includegraphics[height=2.5in, width = 2.1in]{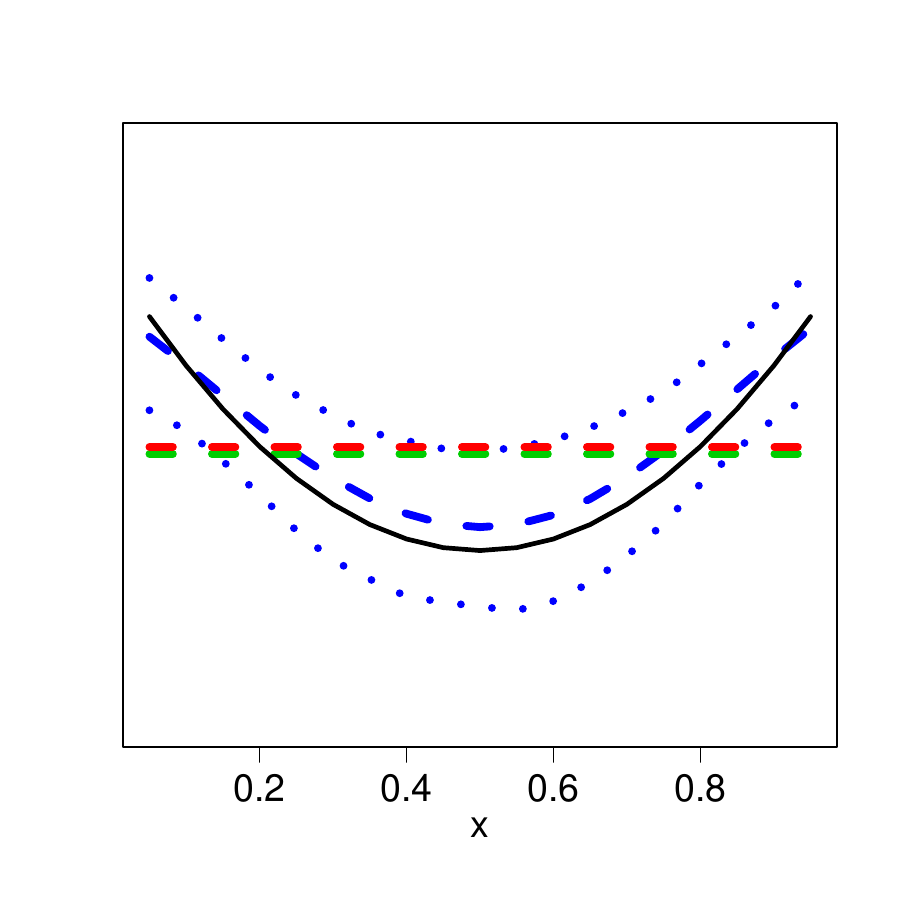}\\
	\end{tabular}
\end{figure}

\newpage
\subsection*{Simulation Results under the Clayton Copula ($n = 100$)}
\label{C}
\begin{table}[H]
	\centering
	\caption{Mean, Integrated Squared Bias ($\text{IBias}^2$) and Integrated Mean Square Error (IMSE) (multiplied by $100$) of the $\hat{S}_{1\mid X}(.)$ at different quantiles over $1000$ Monte Carlo samples under Clayton copula (Low Censoring, $\tau = 0.5$ or $\tau(x)$, $n = 100$).}
	\small	
	\label{tab3}
	\scalebox{0.8}{	\begin{tabular}{p{15mm}ccccccccccccc}
			\\ [-1ex]\hline\\ [-1.5ex]
			&\multirow{2}{*}{p}	&\multicolumn{3}{c}{CGE}&& \multicolumn{3}{c}{CCGE$1$}&& \multicolumn{3}{c}{CCGE$2$}\\
			&&$E(\hat{S}_{1}(\cdot))$& $\text{IBias}^2$& IMSE&&$E(\hat{S}_{1\mid X}(\cdot))$& $\text{IBias}^2$& IMSE &&$E(\hat{S}_{1\mid X}(\cdot))$& $\text{IBias}^2$& IMSE\\
			\hline\\[-1.5ex]
			\multirow{4}{*}{DGM~1}&0.1&0.102& 0.001&  0.107&&0.103&0.001& 0.158&&0.104&0.002&0.257\\
			&0.5&0.502&0.001&0.322&&0.503&0.001&0.493&&0.504&0.002&0.531\\
			&0.9&0.900& 0.000&0.096&&0.900&0.000&0.138&&0.900&0.000&0.138\\[1ex]
			
			\multirow{4}{*}{DGM~2}&0.1&0.124&0.442&0.570&&0.111&0.071&0.320&&0.113&0.075&0.446\\
			&0.5&0.496&0.829&1.127&&0.501&0.098&0.788&&0.502&0.090&0.805\\
			&0.9&0.894&0.078&0.180&&0.899&0.007&0.186&&0.899&0.007&0.186\\[1ex]
			
			\multirow{3}{*}{DGM~3}&0.1&0.122&0.425&0.544&&0.108&0.053&0.282&&0.110&0.055&0.386\\
			&0.5&0.493&0.846&1.126&&0.498&0.087&0.728&&0.498&0.083&0.746\\
			&0.9&0.894&0.078&0.177&&0.899&0.007&0.179&&0.899&0.007&0.181\\
			\hline
	\end{tabular}}
\end{table}

\begin{table}[H]
	\centering
	\caption{Integrated Squared Bias ($\text{IBias}^2$) and Integrated Mean Square Error (IMSE) (multiplied by 100) of the Kendall's tau estimates calculated over $1000$ Monte Carlo samples under Clayton copula (Low Censoring, $\tau = 0.5$ or $\tau(x)$, $n = 100$).}
	\label{tab4}
	\scalebox{0.85}{	\begin{tabular}{cccclccccc}
			\\ [-1ex]	\hline& \\ [-1.5ex]
			& \multicolumn{2}{c}{CGE}&& \multicolumn{2}{c}{CCGE$1$}&& \multicolumn{2}{c}{CCGE$2$}\\
			&$\text{IBias}^2$&  IMSE&& $\text{IBias}^2$& IMSE && $\text{IBias}^2$&IMSE\\
			\hline& \\[-1.5ex]
			DGM~1&0.029 & 0.415 && 0.063&0.471&&0.108& 1.507\\
			DGM~2&0.003&0.357&&0.071&0.489&&0.132&1.540\\    
			DGM~3&1.789&2.164&&1.810&2.239&&0.239&4.111\\
			\hline
	\end{tabular}}
\end{table}

\begin{figure}[H]
	\centering
	\caption{Mean, $5^{\text{th}}$ and $95^{\text{th}}$ quantiles of Kendall's tau estimates under Clayton copula (Low Censoring, $\tau = 0.5$ or $\tau(x)$, $n = 100$) from DGM~1(left), DGM~2 (middle) and DGM~3 (right). Dashed lines represent the Kendall's tau estimates from CGE (red), CCGE$1$ (green), CCGE$2$ (blue); Dotted (blue) and solid (black) lines represents the quantiles of CCGE$2$ and true Kendall's tau.}\vspace{-0.4cm}
	\label{fig3}
	\begin{tabular}{@{\hspace{-2pt}}c@{\hspace{-26pt}}c@{\hspace{-26pt}}c}
		\vspace{-0.6cm}
		\includegraphics[height=2.5in, width = 2.1in]{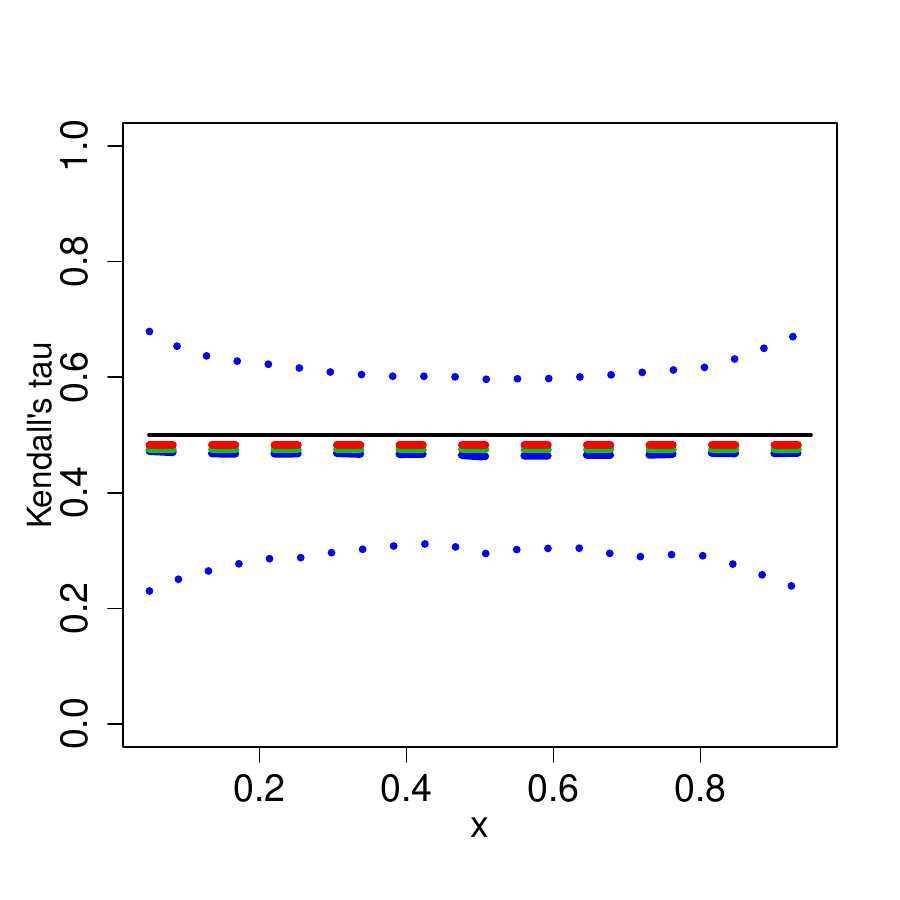} &
		\includegraphics[height=2.5in, width = 2.1in]{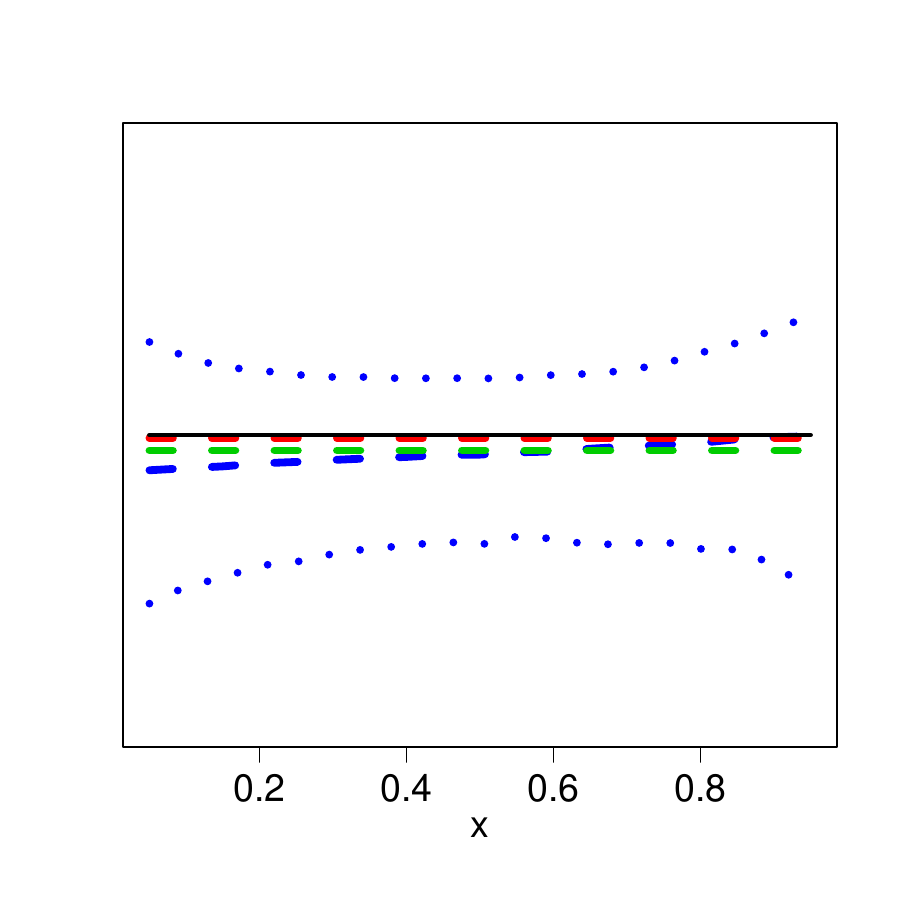} & 
		\includegraphics[height=2.5in, width = 2.1in]{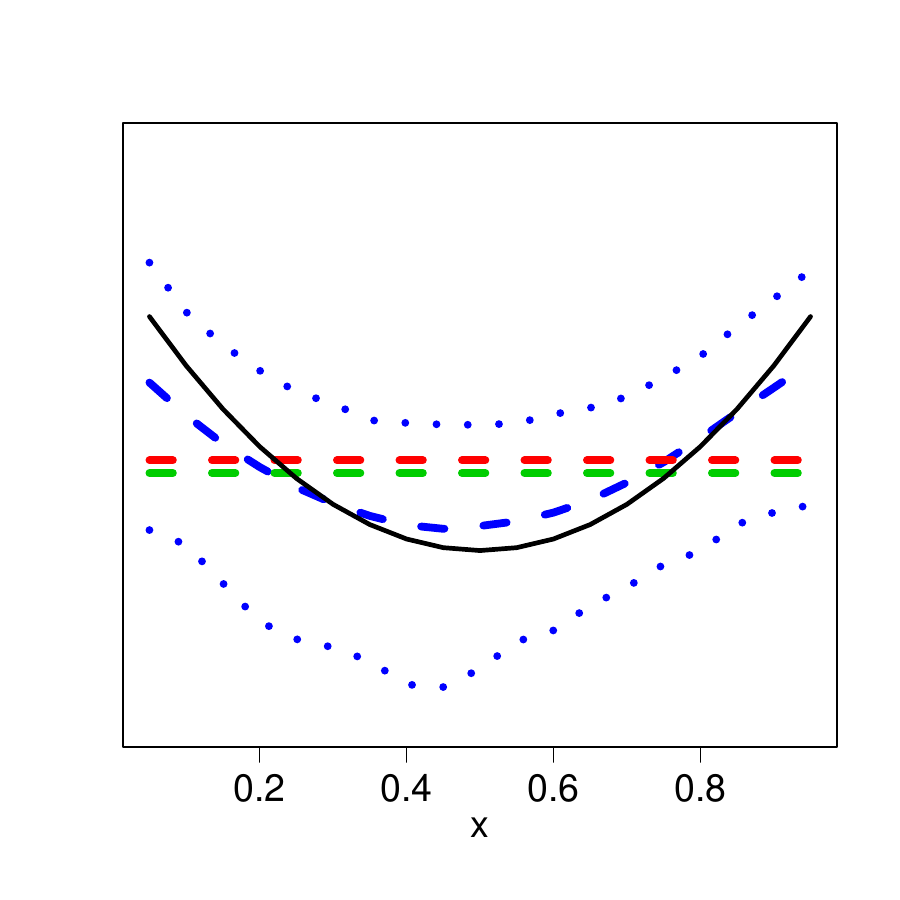}\\
	\end{tabular}
\end{figure}

\clearpage

\subsection*{Simulation Results under the Gumbel Copula ($n = 100$)}
\label{D}
\begin{table}[h]
	\centering
	\caption{Mean, Integrated Squared Bias ($\text{IBias}^2$) and Integrated Mean Square Error (IMSE) (multiplied by 100) of the $\hat{S}_{1\mid X}(.)$ at different quantiles over $1000$ Monte Carlo samples under Gumbel copula (Low Censoring, $\tau = 0.5$ or $\tau(x)$, $n = 100$).}
	\small	
	\label{tab5}
	\scalebox{0.8}{	\begin{tabular}{p{15mm}ccccccccccccc}
			\\ [-1ex]	\hline\\ [-1.5ex]
			&\multirow{2}{*}{p}	&\multicolumn{3}{c}{CGE}&& \multicolumn{3}{c}{CCGE$1$}&& \multicolumn{3}{c}{CCGE$2$}\\
			&&$E(\hat{S}_{1}(\cdot))$& $\text{IBias}^2$& IMSE&&$E(\hat{S}_{1\mid X}(\cdot))$& $\text{IBias}^2$& IMSE &&$E(\hat{S}_{1\mid X}(\cdot))$& $\text{IBias}^2$& IMSE\\
			\hline\\[-1.5ex]
			\multirow{4}{*}{DGM~1}&0.1&0.101&0.000&0.120&&0.101&0.000&0.165&&0.101&0.000&0.167\\
			&0.5&0.500&0.000&0.280&&0.500&0.000&0.423&&0.500&0.000&0.421\\
			&0.9&0.900&0.000&0.092&&0.900&0.000&0.127&&0.900&0.000&0.125\\[1ex]
			
			\multirow{4}{*}{DGM~2}&0.1&0.125&0.449&0.588&&0.109&0.062&0.327&&0.110&0.064&0.327\\
			&0.5&0.494&0.822&1.086&&0.497&0.088&0.691&&0.498&0.088&0.693\\
			&0.9&0.893&0.081&0.178&&0.898&0.007&0.177&&0.899&0.007&0.177\\[1ex]
			\multirow{3}{*}{DGM~3}&0.1&0.123&0.428&0.549&&0.107&0.048&0.281&&0.107&0.053&0.285\\
			&0.5&0.491&0.832&1.087&&0.494&0.085&0.669&&0.493&0.106&0.693\\
			&0.9&0.892&0.083&0.179&&0.897&0.007&0.176&&0.897&0.009&0.180\\
			\hline
	\end{tabular}}
\end{table}

\begin{table}[h]
	\centering
	\caption{Integrated Squared Bias ($\text{IBias}^2$) and Integrated Mean Square Error (IMSE) (multiplied by 100) of the Kendall's tau estimates calculated over $1000$ Monte Carlo samples under Gumbel copula (Low Censoring, $\tau = 0.5$ or $\tau(x)$, $n = 100$).}
	\label{tab6}
	\scalebox{0.85}{	\begin{tabular}{cccclccccc}
			\\ [-1ex]	\hline\\ [-1.5ex]
			& \multicolumn{2}{c}{CGE}&& \multicolumn{2}{c}{CCGE$1$}&& \multicolumn{2}{c}{CCGE$2$}\\
			&$\text{IBias}^2$& IMSE&& $\text{IBias}^2$&IMSE && $\text{IBias}^2$& IMSE\\
			\\ [-1ex]	\hline& \\[-1.5ex]
			DGM~1&0.003&0.374&&0.010&0.373&&0.001&1.152\\
			DGM~2&0.042&0.397&&0.033&0.397&&0.010&1.297\\    
			DGM~3&1.911&2.285&&1.873&2.249&&0.306&3.954\\
			\hline
	\end{tabular}}
\end{table}

\begin{figure}[H]
	\centering
	\caption{Mean, $5^{\text{th}}$ and $95^{\text{th}}$ quantiles of Kendall's tau estimates under Gumbel copula (Low Censoring, $\tau = 0.5$ or $\tau(x)$, $n = 100$) from DGM~1(left), DGM~2 (middle) and DGM~3 (right). Dashed lines represent the Kendall's tau estimates from CGE (red), CCGE$1$ (green), CCGE$2$ (blue); Dotted (blue) and solid (black) lines represents the quantiles of CCGE$2$ and true Kendall's tau.}\vspace{-0.4cm}
	\label{fig5}
	\begin{tabular}{@{\hspace{-2pt}}c@{\hspace{-26pt}}c@{\hspace{-26pt}}c}
		\vspace{-0.6cm}
		\includegraphics[height =2.5in, width =2.1in ]{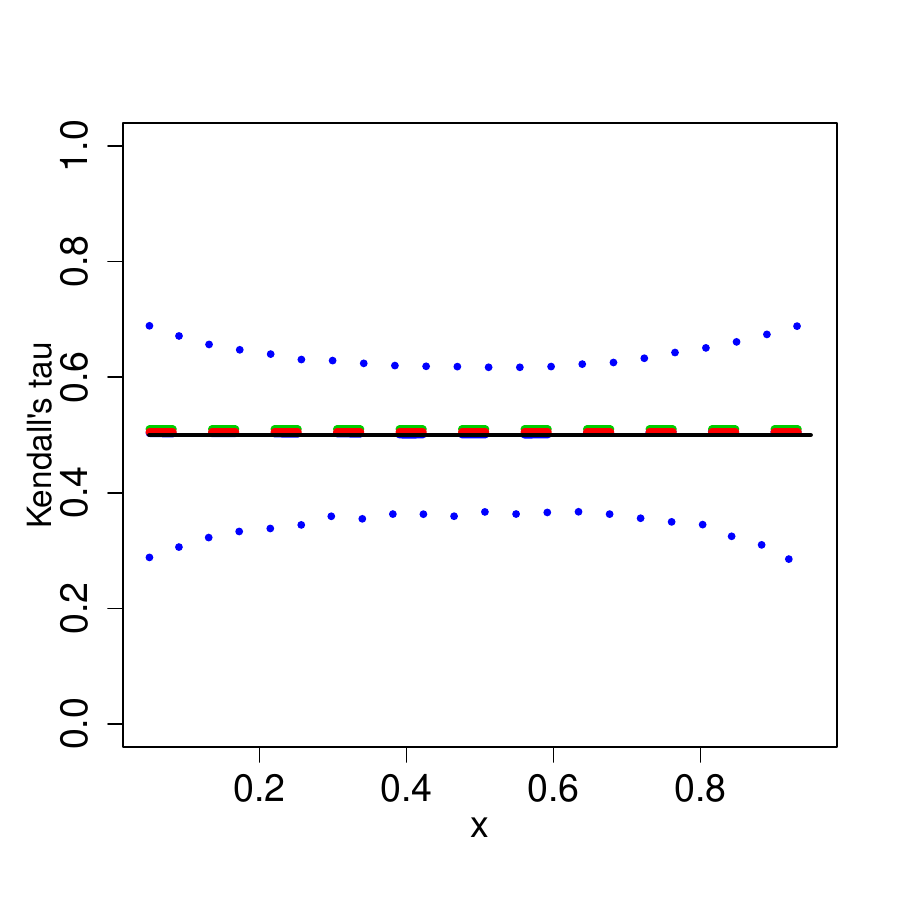} &
		\includegraphics[height =2.5in, width =2.1in ]{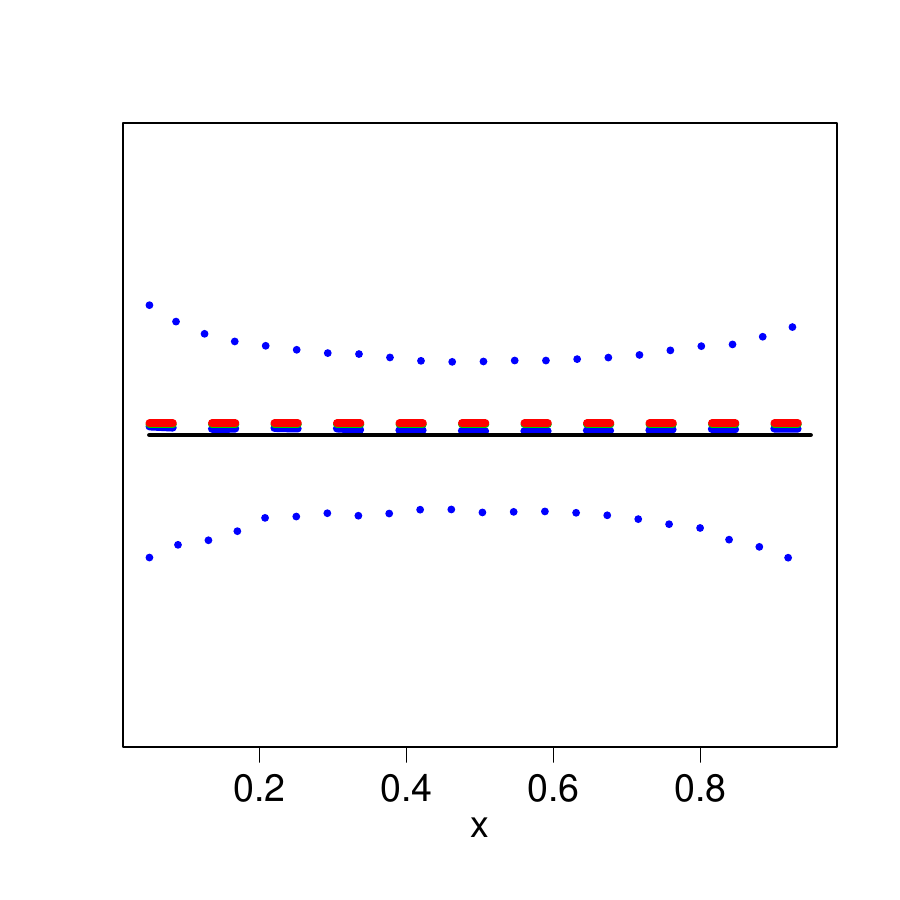} &
		\includegraphics[height =2.5in, width =2.1in ]{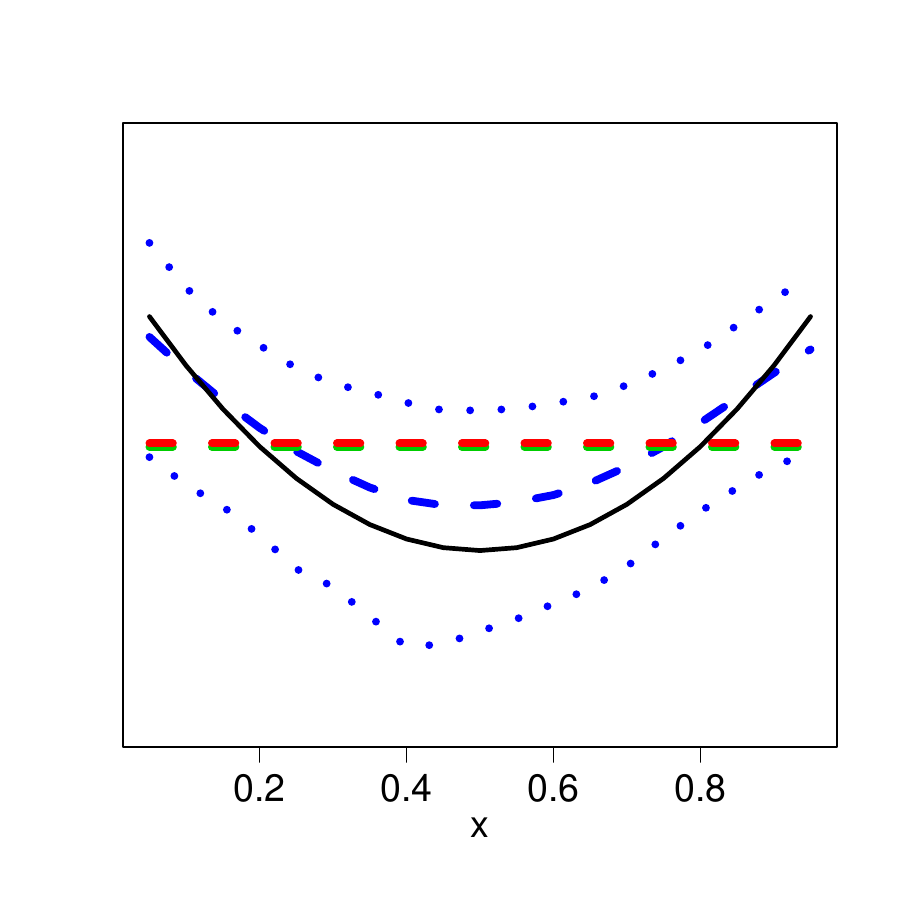}
	\end{tabular}
\end{figure}

\clearpage

\subsection*{Simulation Results under the Frank Copula ($n = 100$, Low Censoring)}
\label{E}
\begin{figure}[H]
	\centering
	\caption{Estimation results and $90\%$ Monte Carlo confidence intervals for the conditional survival function across covariate values at different quantiles ($p = 0.1, 0.5, 0.9$) when data is generated from DGM~1(top row), DGM~2 (middle row) and DGM~3 (bottom row) under Frank copula (Low Censoring, $\tau = 0.5$ or $\tau(x)$, $n = 100$).}\vspace{-0.4cm}
	\label{fig8}
	\begin{tabular}{c@{\hspace{-1pt}}c@{\hspace{-1pt}}c}
		\vspace{-1cm}
		\includegraphics[height=5.7cm, width = 4.8cm]{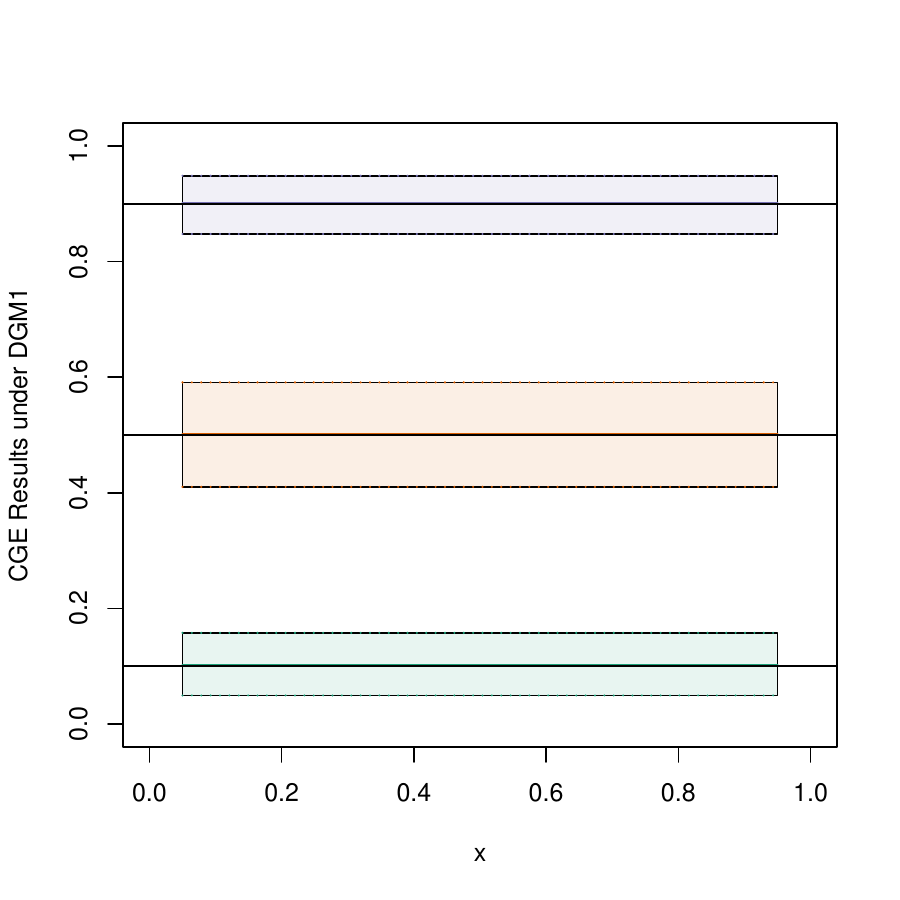} &
		\includegraphics[height=5.7cm, width = 4.8cm]{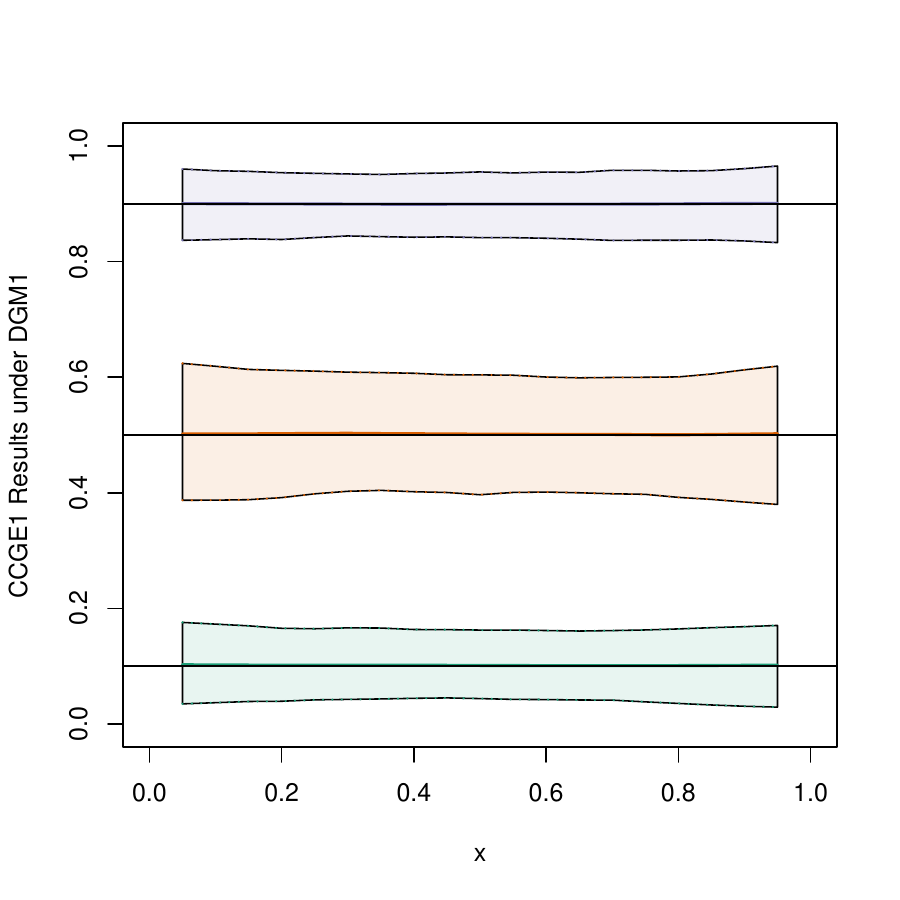} & 
		\includegraphics[height=5.7cm, width = 4.8cm]{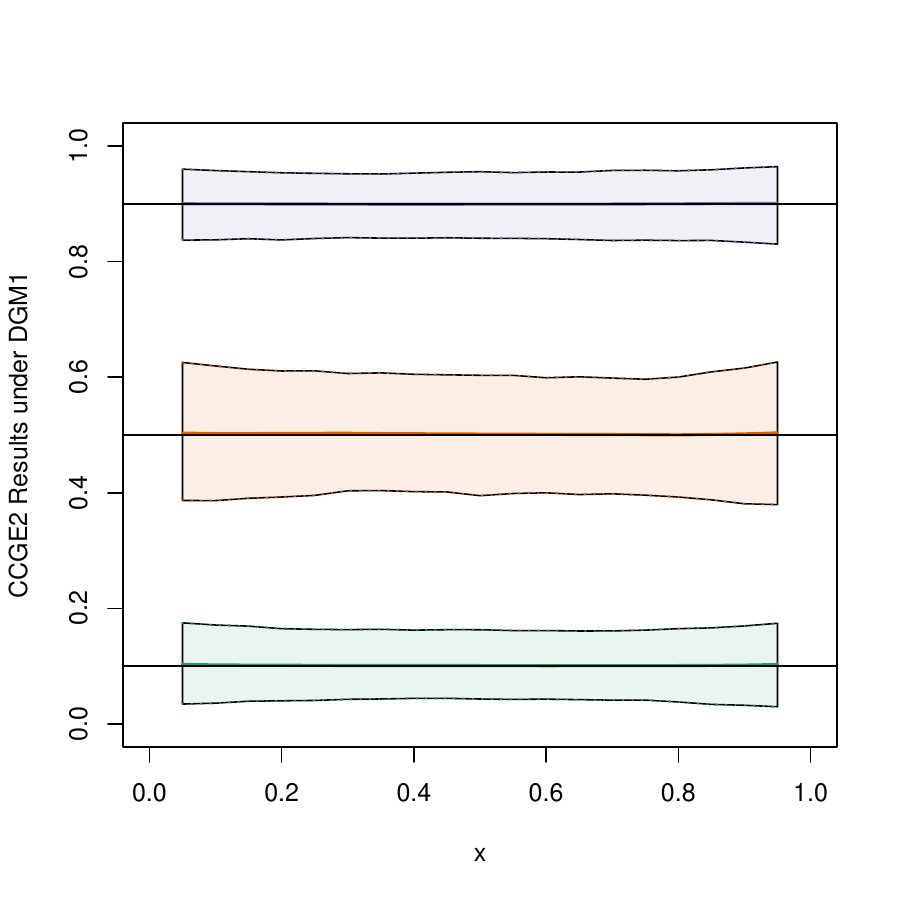}\\	\vspace{-1cm}
		\includegraphics[height=5.7cm, width = 4.8cm]{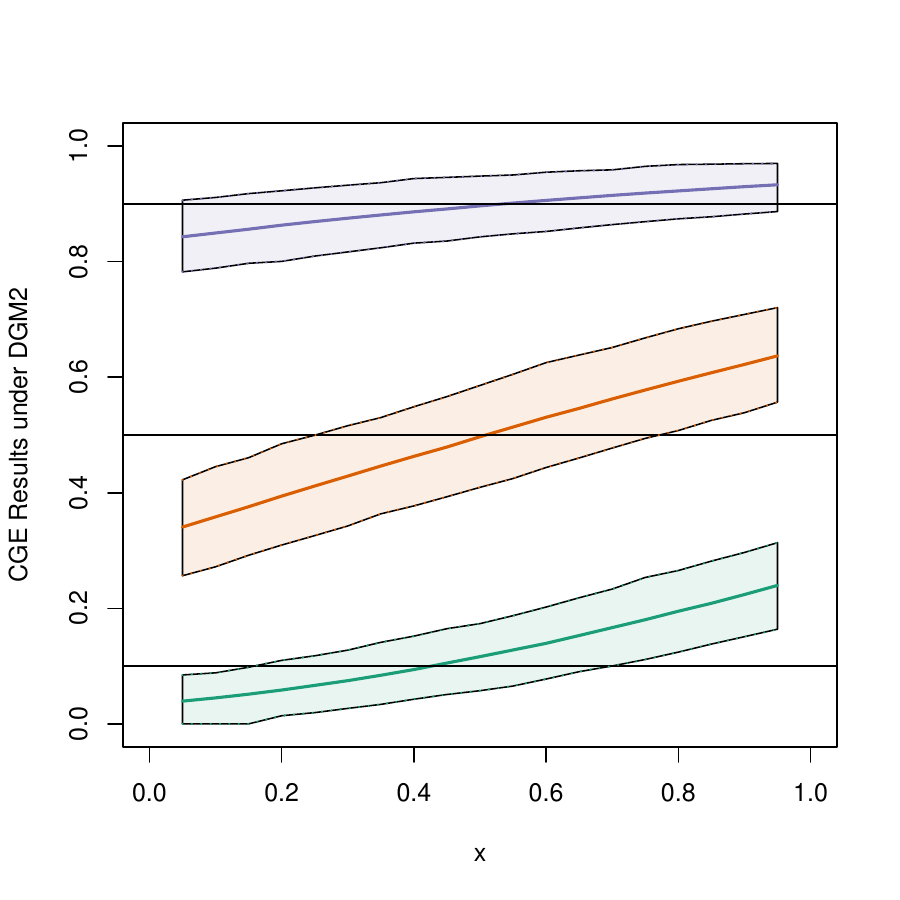} & 
		\includegraphics[height=5.7cm, width = 4.8cm]{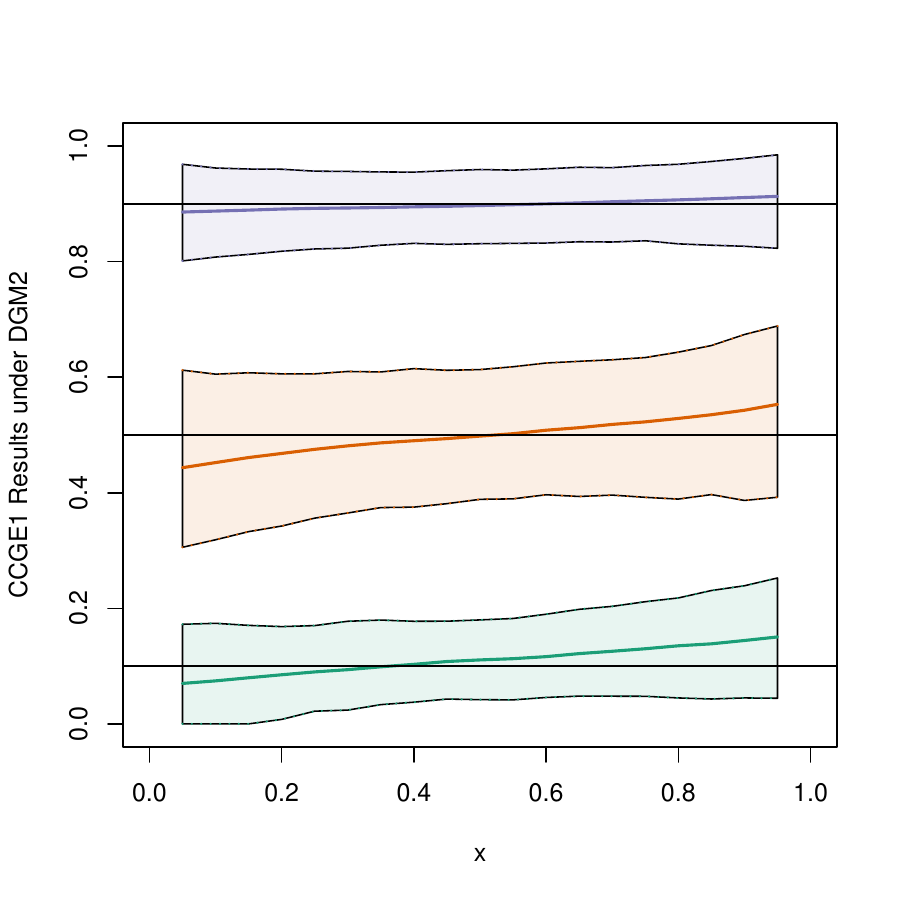} & 
		\includegraphics[height=5.7cm, width = 4.8cm]{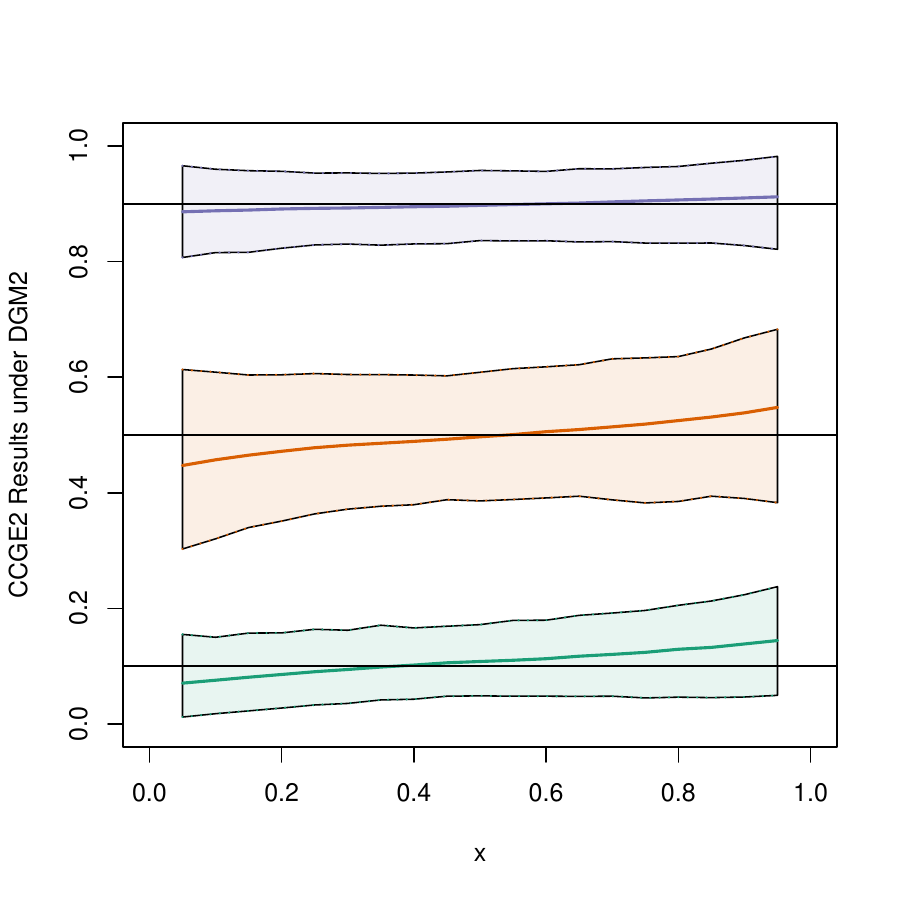}\\\vspace{-1cm}
		\includegraphics[height=5.7cm, width = 4.8cm]{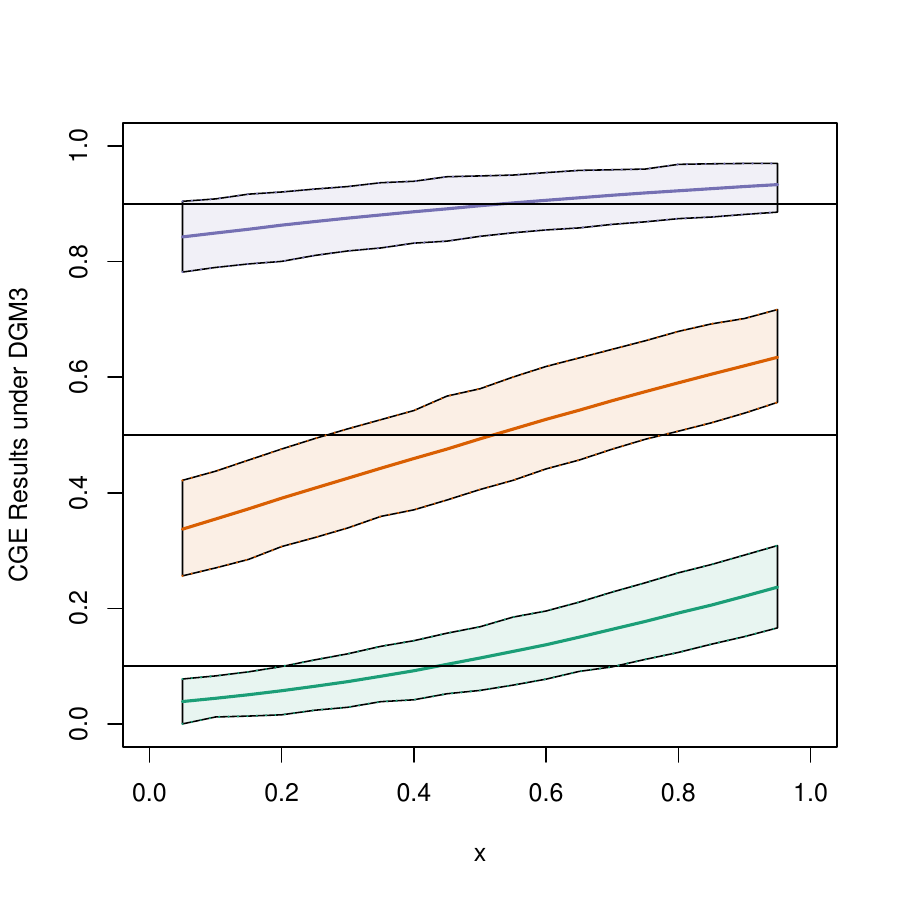} & 
		\includegraphics[height=5.7cm, width = 4.8cm]{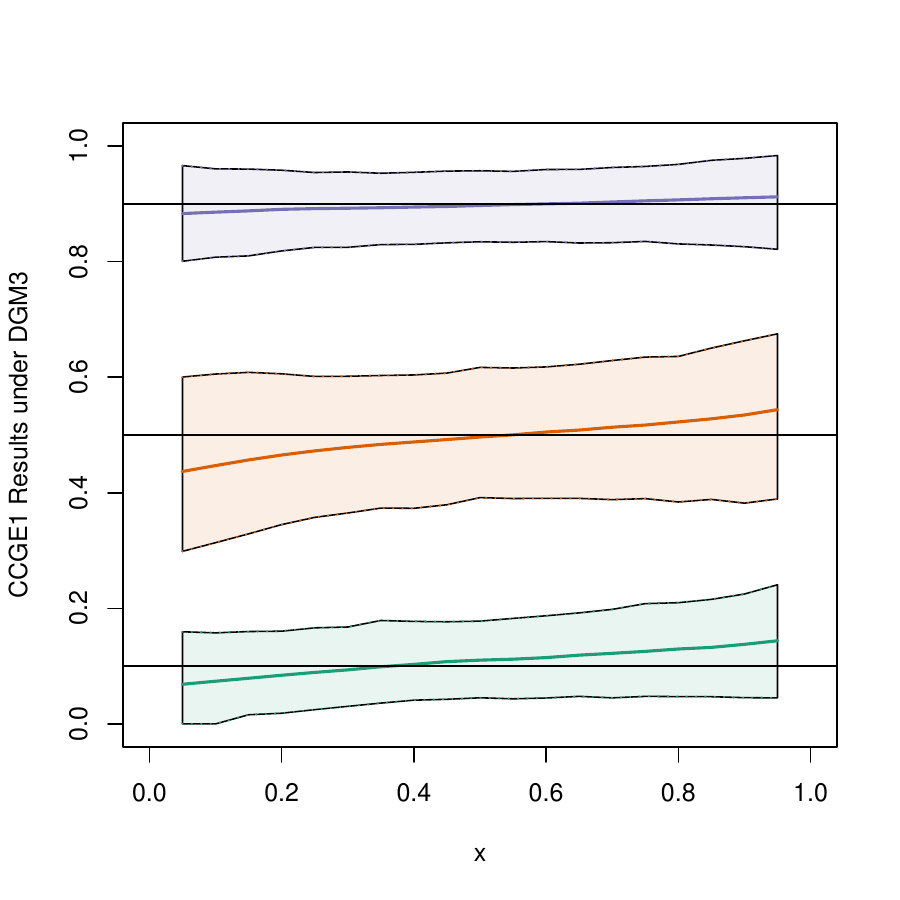} & 
		\includegraphics[height=5.7cm, width = 4.8cm]{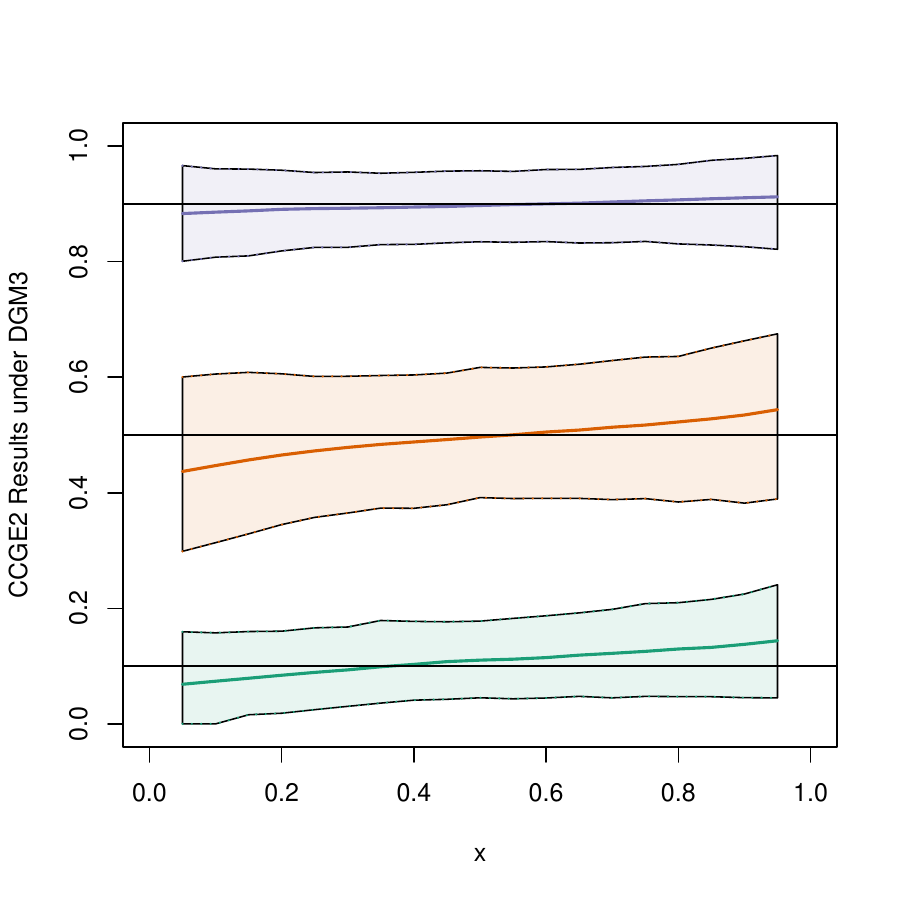}\\
	\end{tabular}
\end{figure}

\clearpage

\subsection*{Selection of the Copula Family (Low Censoring)}
\begin{table}[ht]

\centering

\caption{Number of times out of 1000 simulations the underlying copula family is correctly
identified under each data generating model (DGM) using the maximum copula log-likelihood criterion based on the unconditional (CGE) and conditional copula graphic estimators (CCGE1) assuming a constant dependence parameter.\\}
\scalebox{0.9}{

\begin{tabular}{cc cc cc cc}
\hline
& & \multicolumn{2}{c}{Clayton} & \multicolumn{2}{c}{Gumbel} & \multicolumn{2}{c}{Frank} \\
\cmidrule(r){3-4}\cmidrule(lr){5-6}\cmidrule(lr){7-8}

DGM & $n$ &  CGE & CCGE1 & CGE & CCGE1 & CGE & CCGE1 \\[1ex]
\hline

\multirow{2}{*}{DGM1}
& 100 & 869 & 814 & 932 & 940 & 732 & 660 \\
& 200 & 969 & 953 & 970 & 979 & 937 & 902 \\[1ex]

\multirow{2}{*}{DGM2}
& 100 & 882 & 775 & 906 & 947 & 706 & 589 \\
& 200 & 953 & 912 & 973 & 988 & 908 & 836 \\[1ex]

\multirow{2}{*}{DGM3}
& 100 & 795 & 703 & 906 & 923 & 676 & 614 \\
& 200 & 919 & 864 & 954 & 980 & 864 & 831 \\[1ex]

\hline

\end{tabular}
}
\end{table}

\end{document}